\documentclass[aps,preprint, nofootinbib,secnumarabic,showkeys]{revtex4}
\usepackage{latexsym}
\usepackage{amsfonts}
\usepackage{amsmath}
\usepackage{amssymb}
\usepackage{revsymb}
\usepackage{bm}
\usepackage{graphicx}
\usepackage{varioref,exscale}
\usepackage{subfigure}
\usepackage[vcentermath]{youngtab}
\usepackage{url}

\usepackage{graphicx}
\usepackage{dcolumn}
\usepackage{bm}

\def\beq{\begin{equation}}

\def\eeq{\end{equation}}

\def\beqa{\begin{eqnarray}}

\def\eeqa{\end{eqnarray}}

\begin{document}

\title{{\bf  $\mathbf{SO(10)}$ grand unification in light of recent LHC searches and colored scalars at the TeV-scale}}

\medskip\
 \author{Ufuk Aydemir}
 \email[Email: ]{ufuk.aydemir@physics.uu.se}
\affiliation{Department of Physics and Astronomy, Uppsala University, 
Uppsala 75120, Sweden \vspace{1.0cm}\\
Dedicated to memory of Nam$\textit{\i}$k Kemal Pak (1947-2015).
\vspace{1.0cm}}

\begin{abstract}
\vspace{0.4cm}
We analyze the compatibility of the recent LHC signals and the TeV-scale left-right model(s) in the minimal nonsupersymmetric $SO(10)$ framework. We show that the models in which the Higgs content is selected based on the extended survival hypothesis do not allow the $W_R$ boson to be at the TeV-scale.  By relaxing this conjecture, we investigate various scenarios where a number of  colored-scalars, originated from various Pati-Salam multiplets, are light and whence they survive down to the low energies. Performing a detailed renormalization group analysis with various low-energy Higgs configurations and symmetry breaking chains, while keeping the high energy Higgs content unmodified; we find that, among a number of possibilities, the models which have a light color-triplet scalar, and its combination with a light color-sextet, particularly stand out. Although these models do allow a TeV-scale $W_R$ boson, generating the required value of the gauge coupling $g_R$ at this scale is non-trivial.  
 \end{abstract}
\maketitle



\section{Introduction}
\subsection{Overview}
Following the discovery of the Higgs boson~\cite{atlas:2012gk,cms:2012gu}, the LHC searches have been centered around looking for physics beyond the standard model. The fact that no compelling signals pointing towards new physics have been detected so far has pushed the expectations to the second run of the LHC.

Curiously, ATLAS and CMS recently reported an excess in various search channels in the invariant mass region of 1.8 - 2.0 TeV~\cite{exp,exp2,exp3,exp4,exp5,exp6}, albeit with confidence levels not high enough for calling it a discovery. Nevertheless, in one of the channels, the deviation from the background occurs to be quite noticeable with a local significance of $3.4\sigma$ and a global of $2.5\sigma$~\cite{exp}. It was recently discussed in Ref.~\cite{Brehmer} that these signals can be explained by a heavy gauge boson $W_R$ of the TeV-scale left-right model, with a single coupling $g_R\simeq0.4$. 

It is well known that the left-right (symmetric) model~\cite{Senjanovic:1975rk,Mohapatra:1979ia,Mohapatra:1980yp,Aydemir:2013zua,Aydemir:2014ama} can be incorporated in the SO(10) grand unification scheme~\cite{Mohapatra2,Mohapatra3,Chang:1984qr,Parida:1989an,Deshpande:1992au,Bertolini:2009qj,Babu:2012vc,Awasthi:2013ff,Bandyopadhyay:2015fka,Aydemir:2016qqj}\footnote{For analyses of supersymmetric $SO(10)$ GUT, see Refs.~\cite{Deshpande:1992eu,Majee:2007uv,Parida:2008pu,Parida:2010wq,Dev:2009aw,Fukuyama:2004xs,Fukuyama:2004ps}.}. The gauge group of the model, $SU(2)_L\times SU(2)_R\times U(1)_{B-L}\times SU(3)_C$, can be obtained from the $SO(10)$ group by various symmetry breaking sequences. By breaking $D$-parity at a scale which is different from the breaking scale of $SU(2)_R$~\cite{Mohapatra2,Mohapatra3}, one can also obtain $g_R\neq g_L$ at lower energies, which is required for the compatibility with the recent LHC signals. Note that the value $g_R\simeq0.4$ is different from the value of $g_L$ in the TeV scale.

In this work, we analyze the compatibility of the TeV-scale left-right model embedded in the non-supersymmetric $SO(10)$ framework and the recent LHC signals. First, by performing a detailed renormalization group (RG) analysis, we show that the traditional $SO(10)$ scheme, in which the Higgs content is determined based on the \textit{extended survival hypothesis} (ESH)~\cite{ESH}, does not allow the left-right model to be at the TeV scale. The symmetry breaking scale $M_R$, where the left-right model gauge group is broken into the SM one, turns out to be significantly higher. Recall that the ESH states that at every step of a symmetry breaking chain, the only scalars which survive below the corresponding symmetry breaking scale are the ones which acquire vacuum expectation values (VEV's) at the subsequent levels of the symmetry breaking.  

In order to explore the $SO(10)$ scheme more in depth, we slightly relax the ESH conjecture as effectively and ``economically'' as possible. First of all, we will stay in the minimal picture, by which we mean that we will not include any $SO(10)$ multiplets other than the ones required to begin with. Furthermore, relaxing the ESH in determining the high energy Higgs content does not significantly affect the low energy RG behaviour; therefore in that case, we would have to allow a quite number of large multiplets to survive down to symmetry breaking scales, which would imply excessive amount of fine-tuning in the model. The more effective way to proceed is to allow particles to survive down to $M_R$ from $M_C$, which is the energy scale where the $SU(4)_C$ symmetry is broken\footnote{In Ref.~\cite{Aydemir:2015nfa}, a similar treatment was applied in cases of the partially unified (regular) Pati-Salam model, and the grand unified Pati-Salam model from non-commutative geometry.}. Since it is only single symmetry breaking stage above $M_R$, the fine-tuning is relatively under control\footnote{In the case of models type-II, which will be discussed in the upcoming sections, the situation slightly worsens because of the presence of the energy scale $M_D$ in between $M_R$ and $M_C$.}. Moreover, the colored scalars, which are remnant from breaking of the $SU(4)_C$ gauge group, have potential to change the RG running significantly without being included in large numbers. Therefore, slightly modifying the low energy scalar content by relaxing the ESH generates the possibility to accommodate a TeV-scale $W_R$ boson in the $SO(10)$ framework. As we will see in this work, this is indeed the case. However, the predicted range of values for $g_R(M_R)$ in these models is $g_R\simeq0.47-0.53$, which is above the value given in~\cite{Brehmer}.

\subsection{Status of the recent signals at LHC}

Recently, ATLAS reported on a search for new heavy bosons hadronically decaying into $WW$, $WZ$, or $ZZ$ \cite{exp}. The largest deviation from the background occurs in the $WZ$ channel at around $2$ TeV with a local significance of $3.4\sigma$ and a global of $2.5\sigma$. In addition, both CMS \cite{exp2} and ATLAS \cite{exp3} observe an excess at around $1.8$ TeV in the dijet distributions albeit with low significance ($2.2\sigma$ and $1\sigma$). Moreover, CMS notices an excess, again at around $2$ TeV, both in their search for massive $WH$ production in the $\ell\nu b \overline{b}$ final state \cite{exp4} and in massive resonance production decaying into two SM vector bosons (one of which is leptonically tagged \cite{exp5}),
both of which have lower significance than $2\sigma$. Recently, ATLAS reported in a note on an analysis which combines all diboson searches in all-leptonic, semi-leptonic and all-hadronic final states \cite{exp6}, in which they state that the excesses they observed before in the hadronic channels persist.

In a recent work, it is discussed that the current signals can be explained by a heavy right-handed gauge boson $W_R$ with a single coupling $g_R(M_R)\simeq 0.4$, where $M_R=5$ TeV, in the left-right models with the gauge group $SU(2)_L\times SU(2)_R\times U(1)'$ \cite{Brehmer}. Note this value is different from the value of SM $W_L$ coupling $g_L(5\,\mbox{TeV})\simeq 0.63$ \cite{Agashe:2014kda,ALEPH:2005ab}.

Many other authors have also discussed possible phenomenological consequences of the $W_R$ interpretation~\cite{Aydemir:2015nfa,Mohapatra,Deppisch1,Deppisch2,Deppisch3,Dobrescu:2015qna,Dobrescu:2015yba,Cacciapaglia:2015nga,Fukano:2015uga,Chen:2015xql,Omura:2015nwa,Lane:2015fza,Arnan:2015csa,Coloma:2015una,Saavedra,Heikinheimo,Gluza:2015goa,Hisano:2015gna,Patra:2015bga,Awasthi:2015ota,Sajjad:2015urz}, but we refrain from reviewing them here.

\section{The Left-right model in the minimal SO(10)}

The left-right model of weak interactions is based on the gauge group $SU(2)_L\times SU(2)_R\times U(1)_{B-L}$ with the fermion fields
\begin{equation}
q_L\;=\;\left(\begin{array}{c}u \\d\end{array}\right)_L,\;\;\;\;\;\;q_R\;=\;\left(\begin{array}{c}u \\d\end{array}\right)_R,\;\;\;\;\;\;l_L\;=\;\left(\begin{array}{c}\nu^0 \\e^-\end{array}\right)_L,\;\;\;\;\;\;l_R\;=\;\left(\begin{array}{c}\nu^0 \\e^-\end{array}\right)_R
\end{equation}
with the quantum number assignments
\begin{equation}
(I_L,\;I_R,\;B-L)\;=\;(2,1,\frac{1}{3})\;,\;\;\;\;(1,2,\frac{1}{3})\;,\;\;\;\;(2,1,-1)\;,\;\;\;\;(1,2,-1)\;,
\end{equation}
respectively. The electric charge formula is given by
\begin{equation}
Q\;=\;I_{3L}+I_{3R}+\frac{B-L}{2}\;.
\end{equation}
There are seven gauge bosons in the model, $W_L^i$, $W_R^i$, and $W_{BL}$, $i=1,2,3$, with the gauge couplings $g_L$, $g_R$. and $g_{BL}$, associated with the $SU(2)_L$,  $SU(2)_R$, and $U(1)_{B-L}$ gauge symmetries, respectively.

If the model has the $D$-parity invariance \cite{Maiezza:2010ic}, a $Z_2$ symmetry which maintains a complete symmetry between the left and the right sectors, then the model is called the \textit{left-right symmetric model} (LRSM), and its symmetry group (including the colour sector) is given as $SU(2)_L\times SU(2)_R\times U(1)_{B-L}\times SU(3)_C \times D$ \footnote{Note that the $D$-parity is slightly different than the usual Lorentz parity; the latter does not transform scalars, while the $D$-parity may transform them non-trivially.}. In this case, due to this left-right symmetry, we also have $g_L=g_R$.

If the recent LHC signals are interpreted in the left-right (symmetric) model, they strongly favor that $g_L\neq g_R$ in the TeV-scale \cite{Brehmer}. This can be achieved also from the symmetric case if the the $D$-parity is broken separately at an energy scale ($M_D$) above the TeV-scale, which induces that $g_L\neq g_R$ below the scale $M_D$ since these coupling constants evolve under the influence of different particle contents below this energy scale \cite{Mohapatra2, Mohapatra3}. Then, the symmetry breaking pattern from the gauge group of the left-right model into the Standard Model gauge group is given as\footnote{The symmetry breaking pattern given in Eq.~(\ref{chainbasic}) is not the only option available. Another possible pattern is the one which includes a stage where
$ SU(2)_R \rightarrow U(1)_R$ is followed by $U(1)_R \times U(1)_{B-L} \rightarrow U(1)_Y$ in which an extra scale is assumed. In this case, the gauge bosons $W_R$ and $Z_R$ become massive in different stages. This option is not a subject of this work.}
\begin{eqnarray}
\label{chainbasic}
SU(2)_L\times SU(2)_R\times U(1)_{B-L}\times SU(3)_C\;\xrightarrow{M_R}\;  SU(2)_L\times U(1)_{Y}\times SU(3)_C\;,
\end{eqnarray}
which is followed by the regular breaking into $U(1)_{Q}\times SU(3)_C$ at around $M_Z$.

The Higgs sector, required in order to realize this symmetry breaking pattern, includes $SU(2)_{L,R}$ triplets, $\Delta_{L1} (3,1,2,1)$ and $\Delta_{R1}(1,3,2,1)$, and a bidoublet $\phi(2,2,0,1)$. The triplet $\Delta_{R1}$ breaks $SU(2)_R\times U(1)_{B-L}$ into $U(1)_Y$, while the bidoublet $\phi$ does the same for $SU(2)_L\times U(1)_Y\rightarrow U(1)_Q$, by appropriate VEV's. Note that $\Delta_{L1}$ is introduced only for ensuring the left-right symmetry above $M_D$.\footnote{Here, instead of the $SU(2)$ triplets, the $SU(2)$ doublets $\chi_L(2,1,1,1)$ and $\chi_R(1,2,1,1)$, which originate from the $SO(10)$ multiplet $\mathbf{16}$, can also be used. The advantage of the triplet representation is that it provides a Majorana mass term for the right-handed neutrino.}

The symmetry breaking sequences, required to achieve the symmetry group of the left-right model from $SO(10)$, can be gathered into two groups, depending on whether 
\begin{equation}
\label{condition}
M_D\; \leq \;M_C  \qquad\qquad \mbox{or}\qquad\qquad  M_C\; \leq \;M_D, 
\end{equation}
where $M_C$ is the energy scale at which the $SU(4)_C$ gauge group is broken, while $M_D$ is the $D$-parity breaking scale, mentioned above. Therefore, the most general symmetry breaking sequences are
\begin{eqnarray}
\label{chainsmain}
\mbox{Chain I:}&& \quad SO(10) \;\xrightarrow{M_U}\; G_{224D} \;\xrightarrow{M_D}\; G_{224} \;\xrightarrow{M_C}\;G_{2213}\;\xrightarrow{M_R}\; G_{213} \;\xrightarrow{M_Z}\; G_{13}\;,\nonumber\\
\mbox{Chain II:}&& \quad SO(10) \;\xrightarrow{M_U}\; G_{224D} \;\xrightarrow{M_C}\; G_{2213D} \;\xrightarrow{M_D}\;G_{2213}\;\xrightarrow{M_R}\; G_{213} \;\xrightarrow{M_Z}\; G_{13}\;,\nonumber\\
\end{eqnarray}
where we introduce the notation
\newpage
\begin{eqnarray}
G_{224D}&\equiv&SU(2)_L\times SU(2)_R\times SU(4)_C \times D\;,\cr
G_{224}&\equiv&SU(2)_L\times SU(2)_R\times SU(4)_C\;,\cr
G_{2213D}&\equiv&SU(2)_L\times SU(2)_R\times U(1)_{B-L}\times SU(3)_C \times D\;,\cr
G_{2213}&\equiv&SU(2)_L\times SU(2)_R\times U(1)_{B-L}\times SU(3)_C\;,\cr
G_{213}&\equiv&SU(2)_L\times U(1)_{Y}\times SU(3)_C\;,\cr
G_{13}&\equiv&U(1)_{Q}\times SU(3)_C\;.
\end{eqnarray}
It is also possible to have smaller sequences for each of the conditions given in Eq.~(\ref{condition}), which are 
\begin{eqnarray}
\label{subchains}
\mbox{Chain I-a:} && \;\;\;\;\;\;\;\; SO(10) \;\xrightarrow{M_U=M_D}\; G_{224} \;\xrightarrow{M_C}\;G_{2213}\;\xrightarrow{M_R}\; G_{213} \;\xrightarrow{M_Z}\; G_{13}\;,\nonumber\\
\mbox{Chain I-b:} && \;\;\;\;\;\;\;\; SO(10) \;\xrightarrow{M_U}\; G_{224D} \;\xrightarrow{M_D=M_C}\; G_{2213}\;\xrightarrow{M_R}\; G_{213} \;\xrightarrow{M_Z}\; G_{13}\;,\nonumber\\
\mbox{Chain I-c:} && \;\;\;\;\;\;\;\; SO(10) \;\xrightarrow{M_U=M_D=M_C}\;G_{2213}\; \xrightarrow{M_R}\; G_{213}\;\xrightarrow{M_Z}\; G_{13}\;,\nonumber\\
\mbox{Chain II-a:} && \;\;\;\;\;\;\;\; SO(10) \;\xrightarrow{M_U=M_C}\;G_{2213D}\;\xrightarrow{M_D}\;G_{2213}\; \xrightarrow{M_R}\; G_{213}\;\xrightarrow{M_Z}\; G_{13}\;.
\end{eqnarray}
Note that we ignore the chains with $M_C=M_R$ since we are interested in a TeV-scale $M_R$, and there hasn't been any noticeable signals observed at the LHC regarding a TeV-scale $M_C$.

Our strategy to deal with these symmetry breaking patterns is as follows. It is always possible to start with the most general chain and discover the smaller ones numerically in the process of computation, instead of dealing with each chain separately. However, these two approaches are not always equivalent simply because of the Higgs content chosen to start with in each case. Nevertheless, in the scenarios we explore, they are exactly (numerically) equivalent. Therefore, in this work, we will only consider the models with Chain I and Chain II and cover the subchains, given in Eq.~(\ref{subchains}), numerically in the process.

\section{Set-up}

We would like to see if a TeV-scale left-right model with the required gauge coupling $g_R$ can be accommodated in the $SO(10)$ framework. The most general symmetry breaking sequences, which we will be  concerned with in this work, are given in Eq.~(\ref{chainsmain}).

The ordering of the breaking scales must be strictly maintained in the computations, that is
\begin{eqnarray}
&& M_Z \;\le\; M_R \;\le\; M_C \;\le\; M_D \;\le\; M_U\;\;\;\;\;\mbox{for Chain I}\cr
&&\;\;\;\;\;\;\;\;\;\;\;\;\;\;\;\;\;\;\;\;\;\;\;\;\;\;\;\;\;\mbox{and}\cr
&& M_Z \;\le\; M_R \;\le\; M_D \;\le\; M_C \;\le\; M_U\;\;\;\;\;\mbox{for Chain II}\;.\;
\label{ordering}
\end{eqnarray}
We label the energy intervals in between symmetry breaking scales
starting from $[M_Z,M_R]$ up to $[M_D,M_U]$ for Chain I, and up to $[M_C,M_U]$ for Chain II,  with Roman numerals as:
\begin{eqnarray}
&&\; \mbox{Chain I}\qquad\qquad\qquad\qquad\qquad\qquad\qquad \;\; \mbox{Chain II}\nonumber\\
 \;\;\;\;\;\;\;\;\;\; \mathrm{I}   & \;:\; & [M_Z,\;M_R]\;-\; G_{213}  \;(\mathrm{SM})\;,\; \;\;\;\;\;\;\;\;\;\;\;\;\;\;\;\; \mathrm{I}    \;:\;  [M_Z,\;M_R]\;-\;G_{213} \;(\mathrm{SM}) \;,\cr
\mathrm{II}  & \;:\; & [M_R,\;M_C]\;-\;G_{2213} \;,\;\;\;\;\;\;\;\;\;\;\;\;\;\;\;\;\;\;\;\; \;\;\;\mathrm{II}   \;:\;  [M_R,\;M_D]\;-\;G_{2213}\;,\;\cr
\mathrm{III} & \;:\; & [M_C,\;M_D]\;-\; G_{224}  \;,\;\;\;\;\;\;\;\;\;\;\;\;\;\;\;\;\;\;\;\; \;\;\;\mathrm{III}  \;:\;  [M_D,\;M_C]\;-\; G_{2213D}\;,\;\cr
\mathrm{IV}  & \;:\; & [M_D,\;M_U]\;-\; G_{224D}\;,\;\;\;\;\;\;\;\;\;\;\;\;\;\;\;\;\;\;\;\; \;\mathrm{IV}   \;:\;  [M_C,\;M_U]\;-\; G_{224D}\;.\;
\label{IntervalNumber}
\end{eqnarray}
In several cases, adjacent scales are equal, which collapses the corresponding energy interval
and skips the intermediate step in between. For instance, if $M_D=M_C$ in Chain I, $G_{224D}$ is broken directly into $G_{2213}$, and interval IV will be followed by interval II, skipping interval III. Similarly, when $M_U=M_C$ in Chain II, interval IV does not exist and the RG  running starts from interval III where $G_{2213D}$ is the relevant gauge group.

The boundary/matching conditions we impose on the couplings at the symmetry breaking scales are:
\begin{eqnarray}
M_U & \;:\; & g_L(M_U) \;=\; g_R(M_U) \;=\; g_4(M_U) \;,\label{MUmathcing}\\
M_D & \;:\; &  g_L(M_D) \;=\; g_R(M_D) \;, \vphantom{\Big|} \label{MPmathcing}\\
M_C & \;:\; & \sqrt{\frac{2}{3}}\,g_{BL}(M_C) \;=\; g_3(M_C)=g_4(M_C) \;,\label{MCmatching}\\
M_R & \;:\; & \frac{1}{g_1^2(M_R)} \;=\; \frac{1}{g_R^2(M_R)}+\frac{1}{g_{BL}^2(M_R)}\;,\quad
g_2(M_R)\;=\;g_L(M_R)\;, \label{MRmatching} \\
M_Z & \;:\; & \frac{1}{e^2(M_Z)} \;=\; \frac{1}{g_1^2(M_Z)}+\frac{1}{g_2^2(M_Z)}\;.
\label{MZmatching}
\end{eqnarray}
In the following, we will investigate various scenarios whether it is possible to set $M_R\sim 5\,\mathrm{TeV}$,
while maintaining $M_U$ below the Planck scale.
The IR data which we will keep fixed as boundary conditions to the RG running are
\cite{Agashe:2014kda,ALEPH:2005ab}
\begin{eqnarray}
\alpha(M_Z) & = & 1/127.9\;,\cr
\alpha_s(M_Z) & = & 0.118\;,\cr
\sin^2\theta_W(M_Z) & = & 
0.2312\;,
\label{SMboundary}
\end{eqnarray}
at $M_Z=91.1876\,\mathrm{GeV}$, which translates to
\begin{equation}
g_1(M_Z) \;=\; 0.36\;,\quad
g_2(M_Z) \;=\; 0.65\;,\quad
g_3(M_Z) \;=\; 1.22\;.
\label{MZboundary}
\end{equation}
Note that the coupling constants are all required to remain in the perturbative regime during the
evolution from $M_U$ down to $M_Z$.
\begin{center}
\begin{table}[b]
\caption{Dynkin index $T_i$ for several irreducible representations of $SU(2)$, $SU(3)$, and $SU(4)$.
Note that different normalization conventions are used in the literature.
For example, there is a factor of 2 difference between $T_i$s given in Ref.~\cite{Lindner} and those in Ref.~\cite{Slansky}. We follow the convention of the former. Notice also that there exist two inequivalent 15 dimensional irreducible representations for $SU(3)$.}
{\begin{tabular}{ccccc}
\toprule
\ \ \ Representation\ \ \ \ \ & $\qquad SU(2)\qquad$ & $\qquad SU(3)\qquad$ & $\qquad SU(4)\qquad$ \\
\colrule
$\vphantom{\bigg|}$ 2 &   $\dfrac{1}{2}$ &              $-$ &   $-$ & $\vphantom{\bigg|}$ \\
$\vphantom{\bigg|}$ 3 &                2 &   $\dfrac{1}{2}$ &   $-$ & $\vphantom{\bigg|}$ \\
$\vphantom{\bigg|}$ 4 &                5 &              $-$ &   $\dfrac{1}{2}$ & $\phantom{\bigg|}$ \\
$\vphantom{\bigg|}$ 6 &  $\dfrac{35}{2}$ &   $\dfrac{5}{2}$ &   $1$ & $\vphantom{\bigg|}$ \\
$\vphantom{\bigg|}$ 8 &               42 &              $3$ &   $-$ & $\vphantom{\bigg|}$ \\
$\vphantom{\bigg|}$10 & $\dfrac{165}{2}$ &  $\dfrac{15}{2}$ &   $3$ & $\vphantom{\bigg|}$ \\
$\vphantom{\bigg|}$15 &              280 & $10,\dfrac{35}{2}$ &  4 & $\vphantom{\bigg|}$ \\
\botrule
\end{tabular}
\label{DynkinIndex}}
\end{table}
\end{center}
\vspace{-1.1cm}
\section{One-loop renormalization group running}

For a given particle content, the gauge couplings are evolved according to the 1-loop
RG relation
\begin{eqnarray}
\frac{1}{g_{i}^{2}(M_A)} - \dfrac{1}{g_{i}^2(M_B)}
\;=\; \dfrac{a_i}{8 \pi^2}\ln\dfrac{M_B}{M_A}
\;,
\end{eqnarray}
where the RG coefficients $a_i$ are given by \cite{Jones, Lindner}
\begin{eqnarray}
\label{1loopgeneral}
a_{i}
\;=\; -\frac{11}{3}C_{2}(G_i)
& + & \frac{2}{3}\sum_{R_f} T_i(R_f)\cdot d_1(R_f)\cdots d_n(R_f) \cr
& + & \frac{\eta}{3}\sum_{R_s} T_i(R_s)\cdot d_1(R_s)\cdots d_n(R_s)\;.
\end{eqnarray}
Here, the summation is over irreducible chiral representations of fermions ($R_f$) in the second term and those of scalars ($R_s$) in the third. $\eta=1 \;\mbox{or}\; 1/2$, depending on whether the representation is complex or real, respectively. $C_2(G_i)$ is the quadratic Casimir for the adjoint representation of the group $G_i$,
and $T_i$ is the Dynkin index of each representation. See Table~\ref{DynkinIndex} for the Dynkin indexes of several representations most of which will be useful for our discussion in the following sections. For $U(1)$, $C_2(G)=0$ and
\begin{equation}
\sum_{f,s}T \;=\; \sum_{f,s}\left(\dfrac{Y}{2}\right)^2\;,
\label{U1Dynkin}
\end{equation}
where $Y/2$ is the $U(1)$ charge, the factor of $1/2$ coming from the traditional
normalizations of the hypercharge $d$ and $B-L$ charges.
The $a_i$'s will differ depending on the particle content in each energy interval, which changes every time symmetry breaking occurs. We will distinguish the $a_i$'s in different intervals with the corresponding roman numeral superscript,
cf. Eq.~(\ref{IntervalNumber}).

\section{Models}
\subsection{Models type-I}

We define the models type-I as the models in which $M_D\geqslant M_C$.  Therefore, the relevant most general symmetry breaking sequence is Chain I, which is 
\begin{equation}
SO(10) \;\underset{54}{\xrightarrow{M_U}}\; G_{224D} \;\underset{210}{\xrightarrow{M_D}}\; G_{224} \;\underset{45,\;210}{\xrightarrow{M_C}}G_{2213}\;\underset{126}{\xrightarrow{M_R}}\; G_{213} \;\underset{10}{\xrightarrow{M_Z}}\; G_{13}\;.
\end{equation}
The first stage of the symmetry breaking is realized by a Pati-Salam ($G_{224}$) singlet field acquiring VEV, which is contained in the $SO(10)$ multiplet $\mathbf{54}$ whose decomposition into irreducible representations of $G_{224}$ is given by
\begin{equation}
\mathbf{54}=\left(1,1,1\right)\oplus\left(1,1,20\right)\oplus\left(2,2,6\right)\oplus\left(3,3,1\right)\;.
\end{equation}
Note that the singlet here is even under $D$-parity, which, therefore, remains unbroken at this stage. 

At the second stage, only the $D$-parity is broken, which requires a $G_{224}$ singlet field, odd under $D$-parity. $\mathbf{210}$ contains such a field in its decomposition which is given as
\begin{equation}
\mathbf{210}=\left(1,1,1\right)\oplus\left(2,2,20\right)\oplus\left(3,1,15\right)\oplus\left(1,3,15\right)\oplus\left(2,2,6\right)\oplus\left(1,1,15\right)\;,
\end{equation}
where the required singlet field here is $(1,1,1)_{210}$. $\mathbf{210}$ can also be used to break $G_{224}$ into $G_{2213}$ by the multiplet $(1,1,15)_{210}$. However, note that since $(1,1,15)_{210}$ is even under $D$-parity, it can only be used in the stages where $D$-parity breaking is not required.  If one would like to break the parity together with $SU(4)_C$ as in Chain I-b, given in Eq.~(\ref{subchains}), then one should use $\mathbf{45}$, whose decomposition is given as 
\begin{equation}
\label{decomp3}
\mathbf{45}\;=\;\left(1,1,15\right)\oplus\left(3,1,1\right)\oplus\left(1,3,1\right)\oplus\left(2,2,6\right)\;,
\end{equation} 
where $(1,1,15)_{45}$ is odd under $D$-parity.  Note that $(1,1,15)_{45}$ can also be used for breaking $G_{224}$ into $G_{2213}$ instead of $(1,1,15)_{210}$,  since the parity is not relevant at this stage. We will choose to use $(1,1,15)_{45}\equiv \Sigma (1,1,15)$, since, as mentioned previously, we will be numerically exploring the $M_D=M_C$ case as well in the computations while working out Chain I. Although it does not make a difference numerically to use the either one, $(1,1,15)_{45}$ serves better from the physics perspective.

The breaking of $G_{2213}$ down to $G_{213}$ is accomplished by $(1,3,10)_{126}$, which belongs to $\mathbf{126}$. The decomposition of $\mathbf{126}$ into irreducible representations of $G_{224}$ is given as 
\begin{equation}
\mathbf{126}=\left(1,3,10\right)\oplus\left(3,1,10\right)\oplus\left(2,2,15\right)\oplus\left(1,1,6\right)\;.
\end{equation}
Note that $\mathbf{126}$ provides mass terms for the right-handed and left-handed neutrinos by the multiplets $(1,3,10)_{126}\equiv \Delta_R (1,3,10)$ and $(3,1,10)_{126}\equiv \Delta_L (3,1,10)$, acquiring VEV's; it hence provides both type-I and type-II seesaw mechanism \cite{Bajc:2005zf}.  

Finally, the bidoublet $\phi(2,2,1)$, which contains the required component to realize the electroweak symmetry breaking, i.e. $G_{213} \rightarrow G_{13}$, is found in $\mathbf{10}$ which decomposes into irreducible representations of $G_{224}$ as 
\begin{equation}
\mathbf{10}=\left(2,2,1\right)\oplus\left(1,1,6\right)\;.
\end{equation}
In the following, we will first work out the case where the Higgs content at each energy interval is determined based on the \textit{extended survival hypothesis} (ESH), and then we will proceed to the other models.

\subsubsection{Model I-1: ESH}

Under the ESH, the Higgs sector in the energy interval IV consists of
\begin{equation}
\sigma(1,1,1)\;,\quad\phi(2,2,1)\;,\quad\Delta_R (1,3,10)\;,\quad\Delta_L (3,1,10)\;,\quad\Sigma(1,1,15)\;.
\end{equation}
At the energy scale $M_D$, the symmetry group $G_{224D}$ is broken down to $G_{224}$ by the parity-odd singlet field $\sigma$ acquiring a VEV. According to the ESH, $\Delta_L$ picks a mass at $M_D$ and decouples from the rest.  The remaining fields decompose into irreducible representations of $G_{2213}$ as:
\begin{eqnarray}
\Sigma(1,1,15)
& = & \Sigma_1(1,1,0,1)
\oplus \Sigma_3\left(1,1,\dfrac{4}{3},3\right)
\oplus \Sigma_{\bar{3}}\left(1,1,\dfrac{-4}{3},\bar{3}\right)
\oplus \Sigma_8(1,1,0,8)
\;,\cr
\Delta_R (1,3,10)
& = & \Delta_{R1}(1,3,2,1)
\oplus \Delta_{R3}\left(1,3,\frac{2}{3},3\right)
\oplus \Delta_{R6}\left(1,3,\frac{-2}{3},6\right)
\;, \cr
\phi(2,2,1) & = & \phi(2,2,0,1)\;.\vphantom{\bigg|}
\label{SigmaDeltaphiDecomposition}
\end{eqnarray}
The breaking of $G_{224}$ down to $G_{2213}$ is
realized by the field $\Sigma_1$ acquiring a VEV.
$\Sigma_3$, $\Sigma_{\bar{3}}$, $\Sigma_8$, $\Delta_{R3}$, $\Delta_{R6}$ are all colored-fields, so they do not acquire VEV's in the subsequent steps. Thus, under the ESH, all these fields become heavy at $M_C$ and decouple in the RG equations below $M_C$.

The remaining fields decompose into irreducible representations of $G_{213}$ as:
\begin{eqnarray}
\Delta_{R1}(1,3,2,1) & = & \Delta_{R1}^{0}(1,0,1) \oplus \Delta_{R1}^{+}(1,2,1) \oplus \Delta_{R1}^{++}(1,4,1) \;,\cr
\phi(2,2,0,1) & = & \phi_2(2,1,1) \oplus \phi'_2 (2,-1,1) \;. \vphantom{\bigg|}
\label{Sigma1phiDecomposition}
\end{eqnarray}
The breaking of $G_{2213}$ down to $G_{213}$ is realized by the field $\Delta_{R1}^{0}$, while that of $G_{213}$ down to $G_{13}$ is accomplished by the neutral (diagonal) components of $\phi_2(2,2,0,1)$, acquiring VEVs. The fields $\Delta_{R1}^{+}$ and $\Delta_{R1}^{++}$ are both charged under electromagnetism, so they do not acquire VEV's in the subsequent steps. Thus, these fields become heavy at $M_R$. In addition, only one of the two physical states (which are linear combinations of  $\phi_2$ and $\phi'_2$) remains light while the other picks a mass at $M_R$, unless fine-tuning is applied \cite{finetuning}. The remaining field, the SM Higgs (which can be identified without loss of generality as $ \phi_2(2,1,1)$),  is left to be the only field in the Higgs spectrum below $M_R$.
Thus, the particle content (other than the fermions and gauge bosons) of this model in the energy intervals I through IV are:
\newpage
\begin{eqnarray}
\mathrm{IV} & \;:\; & \sigma(1,1,1)\;,\;\phi(2,2,1)\;,\;\Delta_R (1,3,10)\;,\;\Delta_L (3,1,10)\;,\;\Sigma(1,1,15)\;,\cr
\mathrm{III} & \;:\; & \phi(2,2,1)\;,\;\Delta_R (1,3,10)\;,\;\Sigma(1,1,15)\;,\cr
\mathrm{II}  & \;:\; & \phi(2,2,0,1)\;,\;\Delta_{R1}(1,3,2,1)\;, \vphantom{\bigg|}\cr
\mathrm{I}   & \;:\; & \phi_2(2,1,1)\;.
\end{eqnarray}
The values of the RG coefficients for this Higgs content are listed in Table~\ref{a1}.

Using the relations between the experimentally measured quantities ($\alpha(M_Z)$,
$\alpha_s(M_Z)$, $\sin^2\theta_W(M_Z)$) and the symmetry breaking scales, Eqs.~(\ref{A3}-\ref{A4}), which can be derived by using the one-loop running equations and the boundary/matching conditions, we obtain
\begin{eqnarray}
\label{eqscales1}
2774 &=&
 -46\ln\dfrac{M_U}{M_D}
+36\ln\dfrac{M_D}{M_C}
+57\ln\dfrac{M_C}{M_R}
+109\ln\dfrac{M_R}{M_Z}
\;,\cr
1985 & = &
46\ln\dfrac{M_U}{M_D}
+44\ln\dfrac{M_D}{M_C}
+51\ln\dfrac{M_C}{M_R}
+67\ln\dfrac{M_R}{M_Z}
\;.
\end{eqnarray}
where we also use the RG coefficients given in Table~\ref{a1}.
\begin{table}[t]
\caption{The Higgs content and the RG coefficients in the four energy intervals for Model I-1 where the Higgs selection is made according to the ESH.}
{\begin{tabular}{c|l|l}
\hline
$\vphantom{\Big|}$ Interval & Higgs content & RG coefficients
\\
\hline
$\vphantom{\Biggl|}$   IV
& $\Delta_R (1,3,10),\;\Delta_L (3,1,10),\;\Sigma(1,1,15)$
& $\left( a_{L},a_{R},a_{4}\right)^\mathrm{IV}
 =\left(\dfrac{11}{3},\dfrac{11}{3},-4\right)$
\\
& $\sigma(1,1,1),\;\phi(2,2,1)$ & \\
& &
\\
\hline
$\vphantom{\Biggl|}$ III
& $\phi(2,2,1),\;\Delta_R (1,3,10),\;\Sigma(1,1,15)$
& $\left( a_{L},a_{R},a_{4}\right)^\mathrm{III}
 =\left(-3,\dfrac{11}{3},-7\right)$
\\
\hline
$\vphantom{\Biggl|}$  II
& $\phi(2,2,0,1),\;\Delta_{R1}(1,3,2,1)$
& $\left(a_{L},a_{R},a_{BL},a_3\right)^\mathrm{II}=\left(-3,\dfrac{-7}{3},\dfrac{11}{3},-7\right)$
\\
\hline
$\vphantom{\Biggl|}$   I
& $\phi_2(2,1,1)$
& $\left(a_{1},a_{2},a_{3}\right)^\mathrm{I}=\left(\dfrac{41}{6},\dfrac{-19}{6},-7\right)$
\\
\hline
\end{tabular}}
\label{a1}
\end{table}
To work out the details of Eq.~(\ref{eqscales1}), it is more convenient to work with the common logarithm. Therefore, we make the following definitions.
\begin{eqnarray}
u\;=\;\log_{10}\dfrac{M_U}{\mbox{GeV}}\;,\qquad
d\;=\;\log_{10}\dfrac{M_D}{\mbox{GeV}}\;,\qquad
c\;=\;\log_{10}\dfrac{M_C}{\mbox{GeV}}\;,\qquad
r\;=\;\log_{10}\dfrac{M_R}{\mbox{GeV}}\;.\nonumber\\
\label{xyzDef}
\end{eqnarray}
Then, Eq.~(\ref{eqscales1}) becomes
\begin{eqnarray}
\label{eq1} 
1418 &=&-46 u+82 d + 21 c+ 52r\;,\cr
993 &=& 46 u -2d+ 7 c + 16 r\;.
\end{eqnarray}
Solving the system given in Eq.~(\ref{eq1}) for $u$ and $r$, we obtain
\begin{eqnarray}
\label{eqfloat1a} r &=& 35.5-1.18 d-0.41 c\;,\\
\label{eqfloat1b} u &=& 9.26 +0.45 d-0.01 c\;.
\end{eqnarray}
As can be seen from Eq.~(\ref{eqfloat1a}), the minimum for $r$  is achieved when $d$ and $c$ take their maximum values.  Due to the constraint (\ref{ordering}), the maximum value for $c$ is $d$, and the maximum value for $d$ is $u$. Hence, the minimum value that $r$ is allowed to take can be found from Eq.~(\ref{eqfloat1a}), for $u=d=c$, as
\begin{equation}
\label{min1}
(M_R)_{min}\;:\; \quad M_R\;=\;10^{9.0} \;\mbox{GeV}\;,\;\quad M_U\;=\; M_D\;=\;M_C\;=\; 10^{16.6}\;\mbox{GeV}\;.
\end{equation}
Therefore, the system does not allow that $M_R=5$ TeV.

The maximum value allowed for $r$, again from Eq.~(\ref{eqfloat1a}),  can be found if, this time, $d$ and $c$ take their minimum values, which is $r$. Then for $d=c=r$, we have
\begin{equation}
\label{max1}
(M_R)_{max}\;:\; \quad M_R\;=\; M_C\;=\;M_D\;=\; 10^{13.7} \;\mbox{GeV}\;,\;\quad M_U\;=\; 10^{15.3}\;\mbox{GeV}\;.
\end{equation}
%
\begin{center}
\begin{table}[b]
\caption{The predictions of Model I-1.}
{\begin{tabular}{ccccc}
\toprule
\ \ \ $\;\;M_X$\ \ \ \ \ & $\qquad $ $\log_{10} M_X/GeV$ $\qquad$ \\
\colrule
$\vphantom{\bigg|}$ $M_U$ &   $[15.3, \;16.7]$  \\
$\vphantom{\bigg|}$ $M_D$ &                $[13.7, \;16.7]$  \\
$\vphantom{\bigg|}$ $M_C$ &                $[11.2, \;16.6]$  \\
$\vphantom{\bigg|}$ $M_R$ &  $[9.0, \;13.7]$   \\ \hline
$\vphantom{\bigg|}$ $\alpha_U^{-1}$ &               $[41.0,\;46.4]$  \\
\botrule
\end{tabular}
\label{values1}}
\end{table}
\end{center}

\vspace{-0.93cm}
The maximum value allowed for $M_R$, and the values that $M_U$, $M_D$, and $M_C$ take when $M_R=(M_R)_{max}$ will be the same for the models considered in this work. This is simply because in all of these models $M_R=(M_R)_{max}$ is achieved when $M_R=M_C=M_D$, which collapses the energy interval II and eliminates its effects from the system equations. Since the interval II is the only interval that causes the difference among these models, for the numerical configurations of the ordered quadruple $(M_U,M_D, M_C,M_R)$ (or $(M_U,M_C, M_D,M_R)$ for the models type-II)  which deactivate the interval II, these models will yield identical results. 

\begin{figure}[t]
\subfigure[ \;$(M_R)_{min}:\;\;(r_{min},\;c=d=u)=(9.0,\;16.9)$ ]{\includegraphics[width=7.5cm]{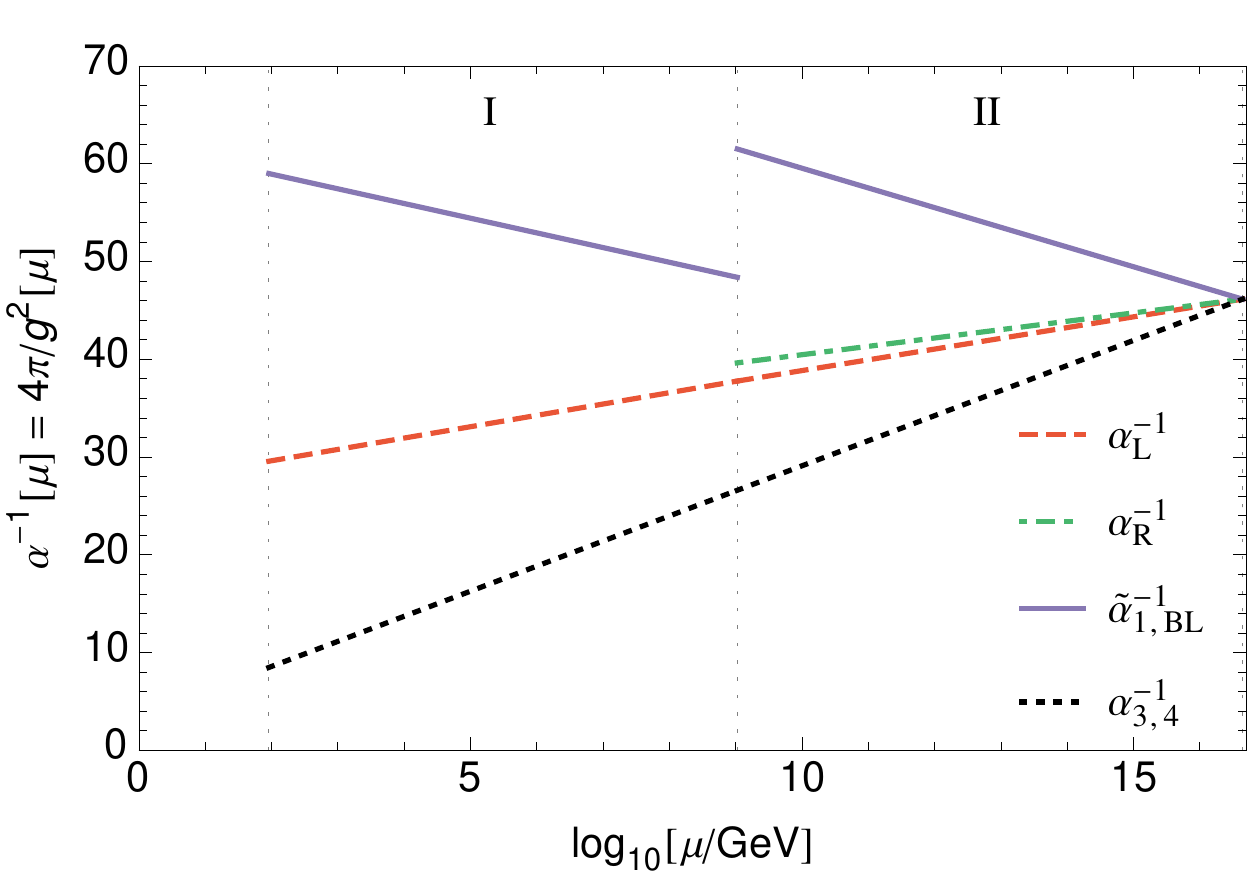}} \hspace{0.4cm} \subfigure[$\;\;(r,c,d,u)=(12.0,\;13.3,\;15.3,\;16.1)$ ]{\includegraphics[width=7.5cm]{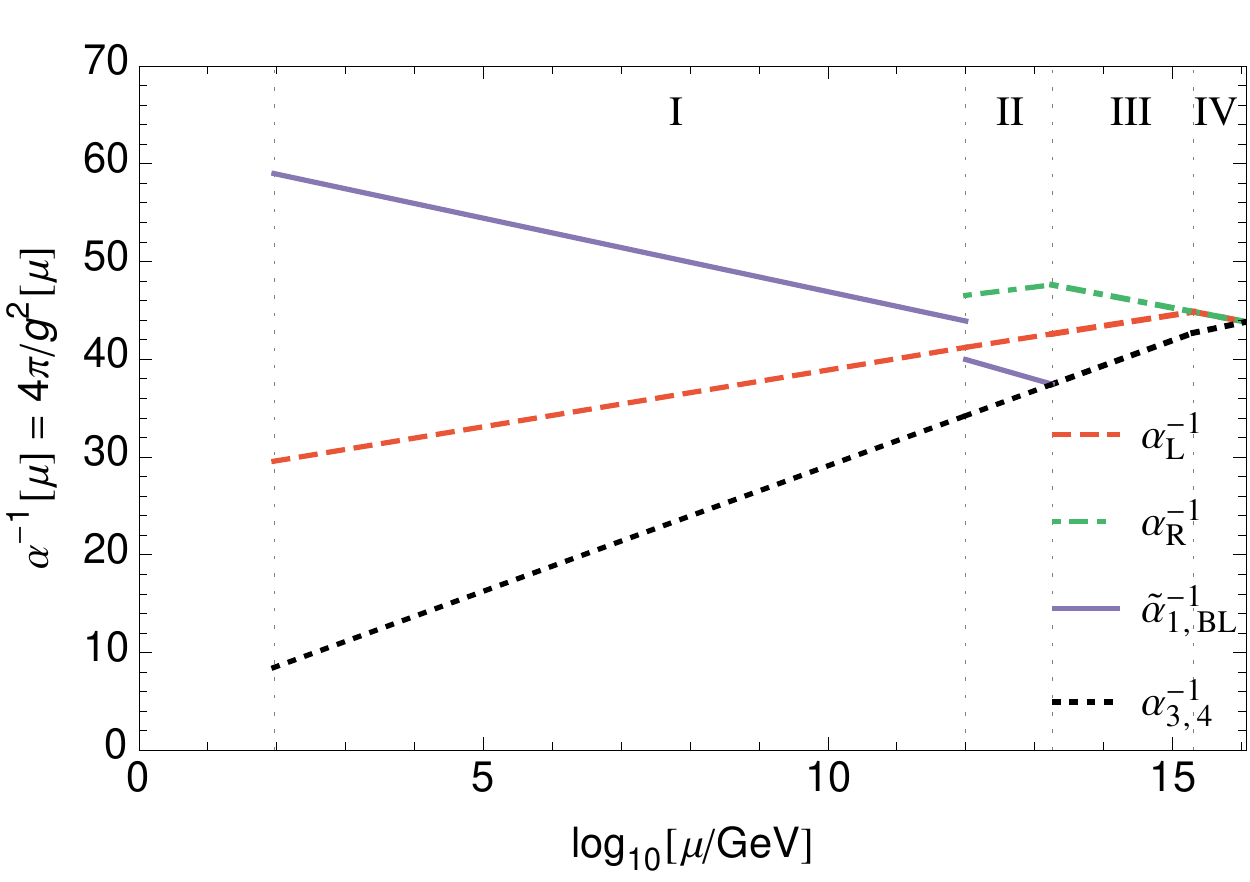}}
\caption{Running of the gauge couplings for Model I-1. The vertical dotted lines from left to right correspond to the
symmetry breaking scales $M_Z$, $M_R$, and $M_C$, and $M_D$, which also indicate the beginning of the energy intervals I, II, III, and IV, respectively. For $\alpha_1^{-1}$ and $\alpha^{-1}_{BL}$, we plot the redefined quantities $\widetilde{\alpha}^{-1}_1\equiv \dfrac{3}{5}\alpha^{-1}_1$ and $\widetilde{\alpha}^{-1}_{BL}\equiv \dfrac{3}{2}\alpha^{-1}_{BL}$. 
The two cases shown are (a) $M_R=(M_R)_{min}=10^{9.0}$ GeV case, and
(b) a random example where we select $M_R=10^{12.0}$ GeV, and among the values now allowed (after fixing $M_R$), we select $M_U=10^{16.1}$ GeV; then, the other values are automatically fixed as $M_D=10^{15.3}$ GeV and $M_C=10^{13.3}$ GeV.
}
\label{RGrunning1}
\end{figure}

Similarly, the interval of values allowed for $M_U$, $M_D$, and $M_C$ can be determined as well; by solving the system equations, given in Eq.~(\ref{eq1}), for the parameter to be determined, while maintaining the ordering of the scales. Additionally, including Eqs.~(\ref{A5}) and (\ref{A6}) into the system, the same procedure can be applied to find the allowed intervals for $\alpha_U$ and $g_R (M_R)$. The results are displayed in Table~\ref{values1}. Note that, throughout this work, we will display the results for $g_R (M_R)$ only if they are relevant to our purpose, i.e.~if the model in question allows $M_R$ to be in the TeV scale. 

Recall that there is another constraint that we impose on our models, which is maintaining $M_U$ below the Planck scale. In this case, however, as can be seen in Table \ref{values1}, the system automatically satisfies this condition.

The running of the coupling constants are displayed in FIG.~\ref{RGrunning1} for two samples of selected values for $(M_U, M_D, M_C, M_R)$.

\subsubsection{Model I-2: A triplet}

We have shown in the previous part that the model in which the Higgs content is determined based on the ESH does not allow $M_R$ to be in the TeV-scale. Now, we would like to relax this conjecture in order to see if it is possible to obtain a different outcome. Recall that we do not change the total particle content of the model which we begin with. Therefore, in that aspect, we are still in the minimal $SO(10)$ framework. The difference now is that we will allow some of the states, in addition to the ones required for the subsequent stages of the symmetry breaking, to be light and survive down to low energies in the RG equation. We do not change the ESH conjecture above $M_C$, where the $SU(4)_C$ is broken. Below this scale, there are only limited number of options available in terms of the sort of particles that can survive down to low energies. The only multiplets that can change the RG behaviour noticeably are the colored scalars originated from $\Delta_R (1,3,10)$ and  $\Sigma (1,1,15)$. As can be seen in their decomposition into irreducible representations of $G_{2213}$, given in Eq.~(\ref{SigmaDeltaphiDecomposition}), there are several color-triplets, a color-sextet, and a color-octet, available for our purpose.  

We begin with investigating whether the color-triplet scalar ($\Delta_{R3}(1,3, 2/3,3)$), which is assumed to be light with a mass of order $M_R$, can enhance the interval of allowed values for $M_R$, found in the previous model, in such a way that it involves TeV-scale values\footnote{Note that color-triplets lead to scalar-induced $d=6$ operators that contribute to the proton decay amplitude. Although these contributions are typically suppressed by small Yukawa couplings, the color-triplets being as light as the TeV-scale can cause a potentially dangerous situation~\cite{Bertolini:2012im}. In that case, a mechanism is required to adequately suppress these interactions, such as the ones proposed in Refs.~\cite{Dvali:1995hp,Rakshit:2003wj}.}. Here, since we have the same picture as before down to $M_C$, there is no change in the energy intervals IV and III in terms of the particle content and the RG coefficients. Below $M_C$, an extra color-triplet Higgs is present down to $M_R$ (interval II) and it is assumed to be decoupled from the rest of the system in the SM interval (interval I), below $M_R$. Therefore, the only changes are in the interval II.  The Higgs content in this interval is given as
\begin{equation}
\phi(2,2,0,1)\;,\quad\Delta_{R1}(1,3,2,1)\;,\quad\Delta_{R3}\left(1,3,\dfrac{2}{3},3\right)\;.
\end{equation}
%

\begin{table}[t]
\caption{ The Higgs content and the corresponding RG coefficients for the models type-I in the energy interval II where the symmetry is $G_{2213}$. Relaxing the ESH leads to different Higgs content and different RG coefficients. Note that the RG coefficients for the other intervals are the same as the ones given in Table \ref{a1}.
}
{\begin{tabular}{c|l|l}
\hline
$\vphantom{\bigg|}$ Models & Higgs content in the energy interval II ($M_C$-$M_R$) & $\left(a_{L},a_{R},a_{BL},a_3\right)^\mathrm{II}$ \\
\hline
$\vphantom{\Biggl|}$  I-1
& $\phi(2,2,0,1),\;\Delta_{R1}(1,3,2,1)$
& $\left(-3,\dfrac{-7}{3},\dfrac{11}{3},-7\right)$
\\
\cline{1-3} I-2
& $\phi(2,2,0,1),\;\Delta_{R1}(1,3,2,1),\;\Delta_{R3}\left(1,3,\dfrac{2}{3},3\right)$
& $\left(-3,\dfrac{-1}{3},4,\dfrac{-13}{2}\right)$ $\vphantom{\Bigg|}$
\\
\cline{1-3} I-3
& $\phi(2,2,0,1),\;\Delta_{R1}(1,3,2,1),\;\Delta_{R6}\left(1,3,\dfrac{-2}{3},6\right)$
& $\left(-3,\dfrac{5}{3},\dfrac{13}{3},\dfrac{-9}{2}\right)$ $\vphantom{\Bigg|}$
\\
\cline{1-3}  I-4
& $\phi(2,2,0,1),\;\Delta_{R1}(1,3,2,1),\;\Sigma_8 (1,1,0,8)$
&   $\left(-3,\dfrac{-7}{3},\dfrac{11}{3},\dfrac{-13}{2}\right)$ $\vphantom{\Bigg|}$
\\
\cline{1-3}  I-5
& $\phi(2,2,0,1),\;\Delta_{R1}(1,3,2,1),\;\Delta_{R3}\left(1,3,\dfrac{2}{3},3\right)\!,\;\Delta_{R6}\left(1,3,\dfrac{-2}{3},6\right)$
& $\left(-3,\dfrac{11}{3},\dfrac{14}{3},-4\right)$ $\vphantom{\Bigg|}$
\\
\cline{1-3}  I-6
& $\phi(2,2,0,1),\;\Delta_{R1}(1,3,2,1),\;\Delta_{R3}\left(1,3,\dfrac{2}{3},3\right)\!,\;\Sigma_{8}\left(1,1,0,8\right)$
&$\left(-3,\dfrac{-1}{3},4,-6\right)$ $\vphantom{\Bigg|}$
\\
\hline
\end{tabular}}
\label{Higgscontent1}
\end{table}

Using Eqs.~(\ref{A3}) and (\ref{A4}) with the new RG coefficients in the interval II, given in Table \ref{Higgscontent1}, we have the following new set of relations.
\begin{eqnarray}
\label{eqscales2}
2774 &=&
 -46\ln\dfrac{M_U}{M_D}
+36\ln\dfrac{M_D}{M_C}
+78\ln\dfrac{M_C}{M_R}
+109\ln\dfrac{M_R}{M_Z}
\;,\nonumber\\
1985 & = &
46\ln\dfrac{M_U}{M_D}
+44\ln\dfrac{M_D}{M_C}
+54\ln\dfrac{M_C}{M_R}
+67\ln\dfrac{M_R}{M_Z}
\;.
\end{eqnarray}
Notice that the only difference between Eq.~(\ref{eqscales1}) and Eq.~(\ref{eqscales2}) is, naturally, the numerical factors in front of $\ln\dfrac{M_C}{M_R}$. In terms of our logarithmic parameters ($u$, $d$, $c$, $r$), defined in Eq.~(\ref{xyzDef}), Eq.~(\ref{eqscales2}) becomes
\begin{eqnarray}
\label{eq2}
1418 &=&-46 u+82 d + 42 c+ 31r\;,\;\cr
 993 &=& 46 u -2d+ 10 c + 13 r\;.\;
\end{eqnarray}
%
%
%
Solving the system given in Eq.~(\ref{eq2}), while maintaining the ordering of the symmetry breaking scales, given in Eq.~(\ref{ordering}), we find
\begin{eqnarray}
\label{minmax1}
(M_R)_{min}\;:\; \qquad M_R=M_Z\;,\qquad  \frac{M_U}{\mbox{GeV}}&=&\left[10^{18.0},\; 10^{18.2}\right]\;,
\quad \frac{M_D}{\mbox{GeV}}\;=\;\left[10^{17.6},\; 10^{18.2}\right]\;\nonumber\\
 \frac{M_C}{\mbox{GeV}}&=&\left[10^{16.7},\; 10^{17.6}\right]\;.
\end{eqnarray}
Unlike the previous model, the system equations in this case cannot pin the values that $M_U$, $M_D$, and $M_C$ take when $M_R=(M_R)_{min}$ to single values; instead, they are given in terms of intervals. When $M_R$ is allowed to float between its minimum and maximum, these intervals naturally become wider. The results for this case are given in Table~\ref{values2}a. The maximum value allowed for $M_R$, and the ones that $M_U$, $M_D$, and $M_C$ take when $M_R=(M_R)_{max}$,  are the same as the ones in the previous model, given in Eq.~(\ref{max1}).

As a result, the system allows a TeV-scale $M_R$. The values $M_U$, $M_D$, and $M_C$ can take, when $M_R=5$ TeV, are given in Table~\ref{values2}b. Since the main prediction we are interested in is the value of $g_R (M_R)$, using Eqs.~(\ref{A5}) and (\ref{eq1}), we obtain


\begin{table}[t]
  \centering 
  \caption{The predictions of Model I-2 for cases where $M_R$ is allowed to float and where it is fixed to $5$ TeV.}
  \label{values2}
  \begin{tabular}{cc}
{\begin{tabular}{ccccc}
\toprule
\ \ \ $\;\;M_X$\ \ \ \ \ & $\qquad $ $\log_{10} M_X/GeV$ $\qquad$ \\
\colrule
$\vphantom{\bigg|}$ $M_U$ &   $[15.3, \;18.2]$  \\
$\vphantom{\bigg|}$ $M_D$ &                $[13.7, \;18.2]$  \\
$\vphantom{\bigg|}$ $M_C$ &                $[11.2, \;17.6]$  \\
$\vphantom{\bigg|}$ $M_R$  &  $[M_Z,\;13.7]$  \\ \hline
$\vphantom{\bigg|}$ $\alpha_U^{-1}$ &               $[41.0,\;47.4]$  \\ \hline
$\vphantom{\bigg|}$ $ g_R(M_R) $ &               $[0.48,\;0.54]$  \\
\botrule
\end{tabular}
}\qquad&\qquad
 {\begin{tabular}{ccccc}
  \toprule
\ \ \ $\;\;M_X$\ \ \ \ \ & $\qquad $ $\log_{10} M_X/GeV$ $\qquad$ \\
\colrule
$\vphantom{\bigg|}$ $M_U$ &   $[17.6, \;17.9]$  \\
$\vphantom{\bigg|}$ $M_D$ &                $[17.0, \;17.9]$  \\
$\vphantom{\bigg|}$ $M_C$ &                $[15.7, \;17.0]$  \\
$\vphantom{\bigg|}$ $M_R$                  &  $5$ TeV   \\ \hline
$\vphantom{\bigg|}$ $\alpha_U^{-1}$ &               $[45.5,\;47.2]$  \\ \hline
$\vphantom{\bigg|}$ $ g_R(M_R) $ &               $[0.51,\;0.53]$  \\
\botrule
\end{tabular}
}\\
(a) $M_R$ floating:  & (b) $M_R$ fixed:
\end{tabular}
\end{table}

\begin{equation}
\frac{1}{g_R^2 (M_R)}\;=\;-13.85+0.99u\;,\;
\end{equation}
which, together with the maximum and minimum values allowed for $u$, yields
\begin{equation}
0.51\leq g_R(M_R) \leq0.53
\end{equation}
for $M_R=5$ TeV. The running of the coupling constants for this case is given in FIG.~\ref{RGrunning2} (a). 

\subsubsection{Model I-3:  A sextet }
In this model, we assume that only the color-sextet component ($\Delta_{R6}$) of $\Delta_R (1,3,10)$ is light and survives down to the mass scale $M_R$ (inteval II). Then, the Higgs content in the interval II becomes
\begin{equation}
\phi(2,2,0,1)\;,\quad\Delta_{R1}(1,3,2,1)\;,\quad\Delta_{R6}\left(1,3,\dfrac{-2}{3},6\right)\;.
\end{equation}
The corresponding RG coefficients for this interval are given in Table \ref{Higgscontent1}, and the ones for the other intervals are the same as before, given in Table \ref{a1}. Numerically, we have
\begin{eqnarray}
\label{eq3}
1418 &=&-46 u+82 d + 63 c+ 10r\;,\;\cr
 993 &=& 46 u -2d+ c +22 r\;.\;
\end{eqnarray}
%
\begin{center}
\begin{table}[b]
\caption{The predictions of Model I-3. The underlined value in the first row implies that we employ the condition of maintaining $M_U$ below the Planck mass.}
{\begin{tabular}{ccccc}
\toprule
\ \ \ $M_X$\ \ \ \ \ & $\qquad $ $\log_{10} M_X/GeV$ $\qquad$ \\
\colrule
$\vphantom{\bigg|}$ $M_U$ &   $[15.3, \;\underline{19.0}]$  \\
$\vphantom{\bigg|}$ $M_D$ &                $[13.7, \;19.0]$  \\
$\vphantom{\bigg|}$ $M_C$ &                $[10.6, \;15.4]$  \\
$\vphantom{\bigg|}$ $M_R$ &  $[6.1, \;13.7]$   \\ \hline
$\vphantom{\bigg|}$ $\alpha_U^{-1}$ &               $[39.7,\;48.6]$  \\
\botrule
\end{tabular}
\label{valuesI3}}
\end{table}
\end{center}

\vspace{-1cm}
Solving these equations for $r$ and $u$ while maintaining the ordering of the symmetry breaking scales, we find that $(M_R)_{min}= M_Z$ is achieved when $M_U/\mbox{GeV}= [10^{21.0},\;10^{21.4}]$. Setting $M_R=5$ TeV yields
\begin{eqnarray}
\label{minbad}
M_R\;=\; 5 \mbox{ TeV}\;, \qquad  \frac{M_U}{\mbox{GeV}}&=&\left[10^{20.2},\; 10^{20.5}\right]\;,
\quad \frac{M_D}{\mbox{GeV}}\;=\;\left[10^{15.9},\; 10^{20.5}\right]\;\nonumber\\
 \frac{M_C}{\mbox{GeV}}&=&\left[10^{10.2},\; 10^{15.9}\right]\;, \quad g_R(M_R)\;=\;\left[0.43,\;0.49\right] \;.
 \end{eqnarray}
Note that $M_U$ exceeds the Planck scale, whereas we would like to keep it below this scale. If we employ this condition, we obtain
\begin{eqnarray}
\label{min2}
(M_R)_{min}:\qquad M_R\;=\; 10^{6.1} \mbox{ GeV}\;,\quad M_U\;=\;10^{19.0}\mbox{ GeV}\;,\quad M_D\;=\;M_C\;=\;  10^{15.4}\;\mbox{GeV}\;.\;\nonumber\\
\end{eqnarray}
which, obviously, excludes $M_R=5$ TeV. The rest of the results for this case are displayed in Table~\ref{valuesI3}, and the running of the coupling constants is given in FIG.~\ref{RGrunning2} (b), for a sample of values of the symmetry breaking scales.

\subsubsection{Model I-4:  An octet}
In this model, we investigate the case of the color-octet $\Sigma_8 (1,1,0,8)$, which is a part of the multiplet $\Sigma(1,1,15)$, surviving in the energy interval II ($M_C-M_R$). The Higgs content in the interval II is then given as
\begin{equation}
\phi(2,2,0,1)\;,\quad\Delta_{R1}(1,3,2,1)\;,\quad\Sigma_{8}\left(1,1,0,8\right)\;.
\end{equation}
Using Eqs.~(\ref{A3}-\ref{A4}) and the corresponding RG coefficients given in Table~\ref{a1}, in terms of the definitions given in Eq.~(\ref{xyzDef}), we obtain
\begin{eqnarray}
\label{eq4}
1418 &=&-46 u+82 d + 21 c+ 52r\;,\;\cr
 993 &=& 46 u -2d+3 c + 20 r\;.\;
\end{eqnarray}
%
\begin{center}
\begin{table}[b]
\caption{The predictions of Model I-4. Note that the system itself maintains $M_U$ below the Planck scale, unlike the one in the previous model.}
{\begin{tabular}{ccccc}
\toprule
\ \ \ $M_X$\ \ \ \ \ & $\qquad $ $\log_{10} M_X/GeV$ $\qquad$ \\
\colrule
$\vphantom{\bigg|}$ $M_U$ &   $[15.3, \;17.9]$  \\
$\vphantom{\bigg|}$ $M_D$ &                $[13.7, \;17.9]$  \\
$\vphantom{\bigg|}$ $M_C$ &                $[11.2, \;17.9]$  \\
$\vphantom{\bigg|}$ $M_R$ &  $[7.7, \;13.7]$   \\ \hline
$\vphantom{\bigg|}$ $\alpha_U^{-1}$ &               $[41.0,\;47.4]$  \\
\botrule
\end{tabular}
\label{valuesI4}}
\end{table}
\end{center}
\vspace{-1cm}
Solving these equations while maintaining the ordering of the scales, 
the minimum possible value for $M_R$ is found as
\begin{equation}
\label{min3}
(M_R)_{min}\;:\;\qquad M_R\;=\;10^{7.7} \mbox{ GeV}\;,\;\quad M_U\;=\;M_D\;=\;M_C\;=\; 10^{17.9}\mbox{ GeV}\;,
\end{equation}
while the ordered quadruple $(M_U,M_D,M_C,M_R)$ for $M_R=(M_R)_{max}$ is the same as before, given in Eq.~(\ref{max1}).
\begin{figure}
 \subfigure[$\;\;(r,c,d,u)=(5 \mbox{ TeV},\;15.7,\;17.9,\;17.9)$ ]{\includegraphics[width=6.7cm]{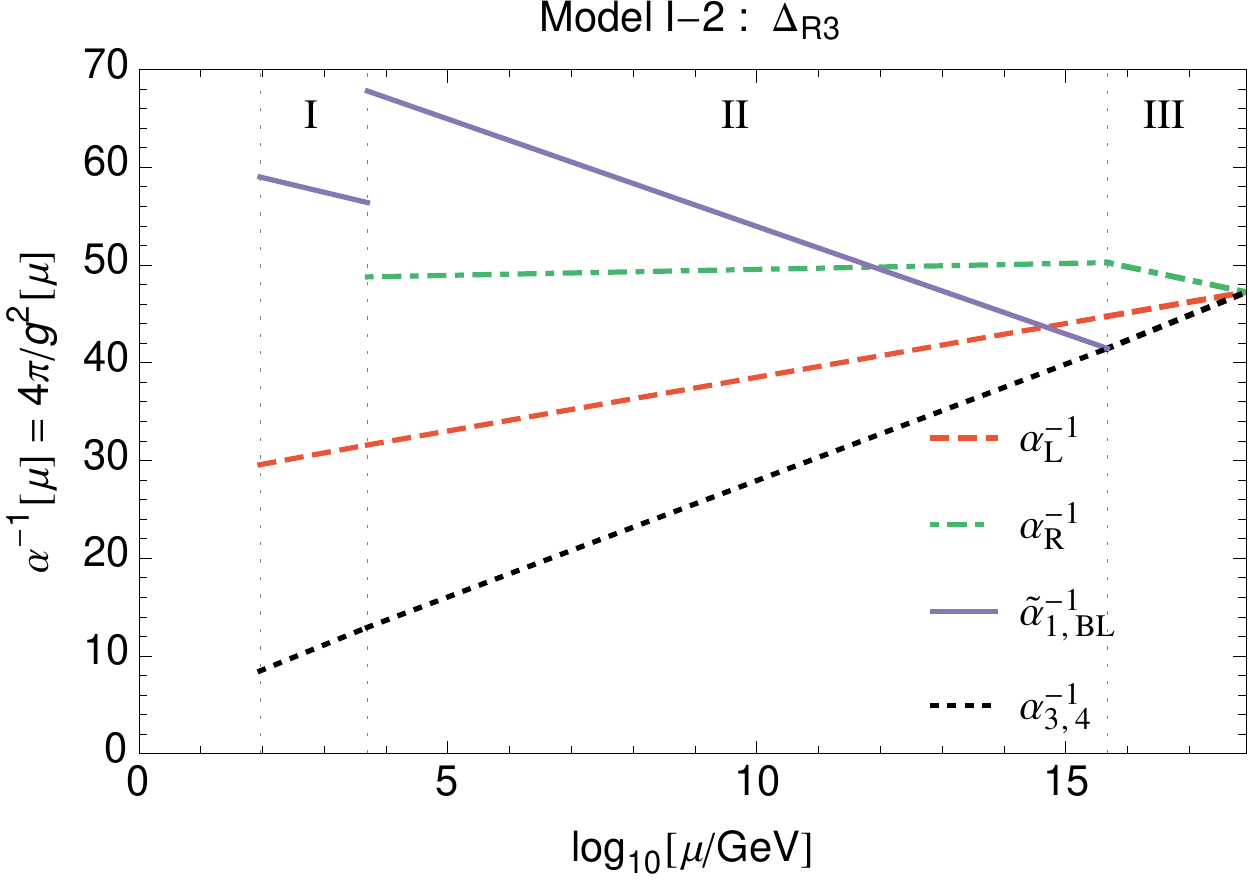}}\hspace{0.4cm}
  \subfigure[$\;\;(r,c,d,u)=(10.0,\;13.3,\;15.5,\;17.2)$ ]{\includegraphics[width=6.7cm]{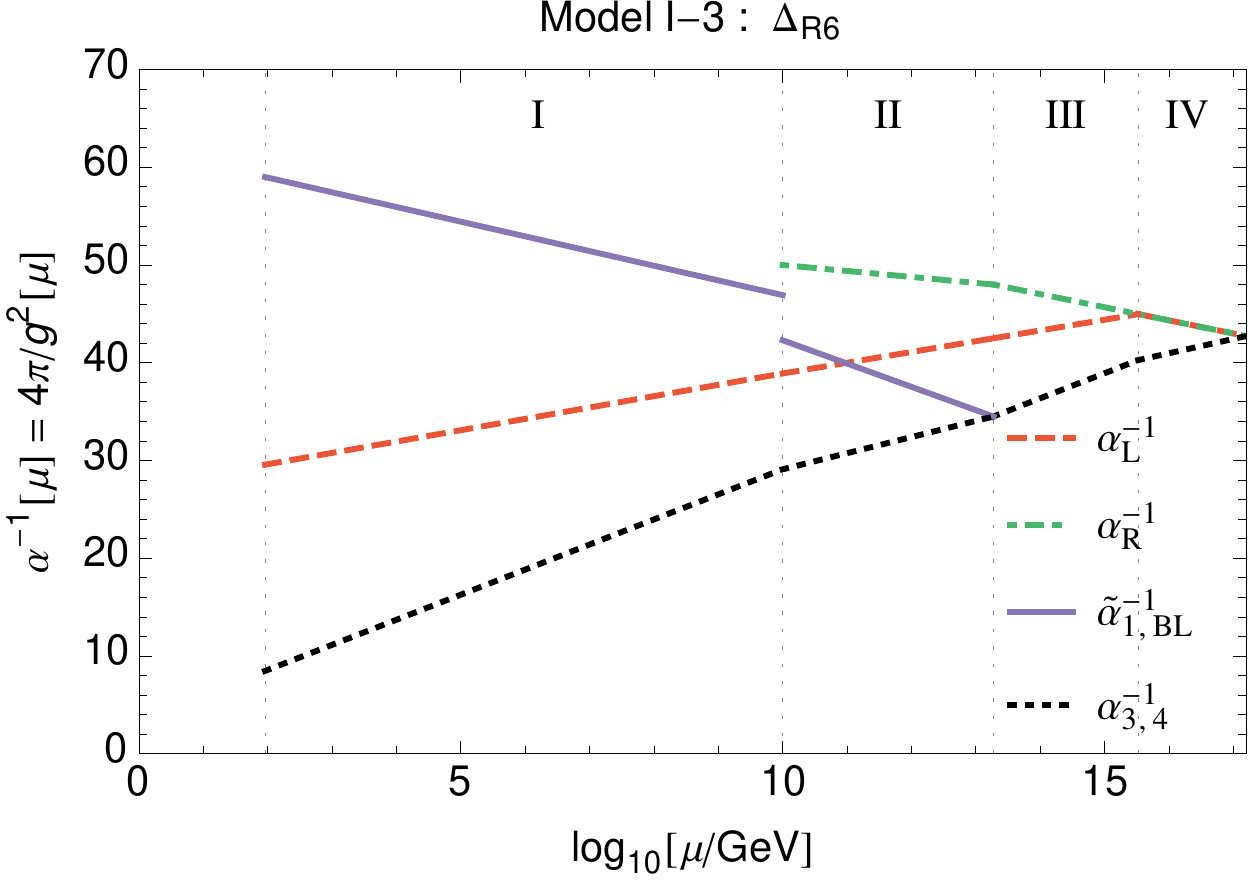}}
  \subfigure[$\;\;(r,c,d,u)=(9.0,\;15.4,\;17.4,\;17.4)$ ]{\includegraphics[width=6.7cm]{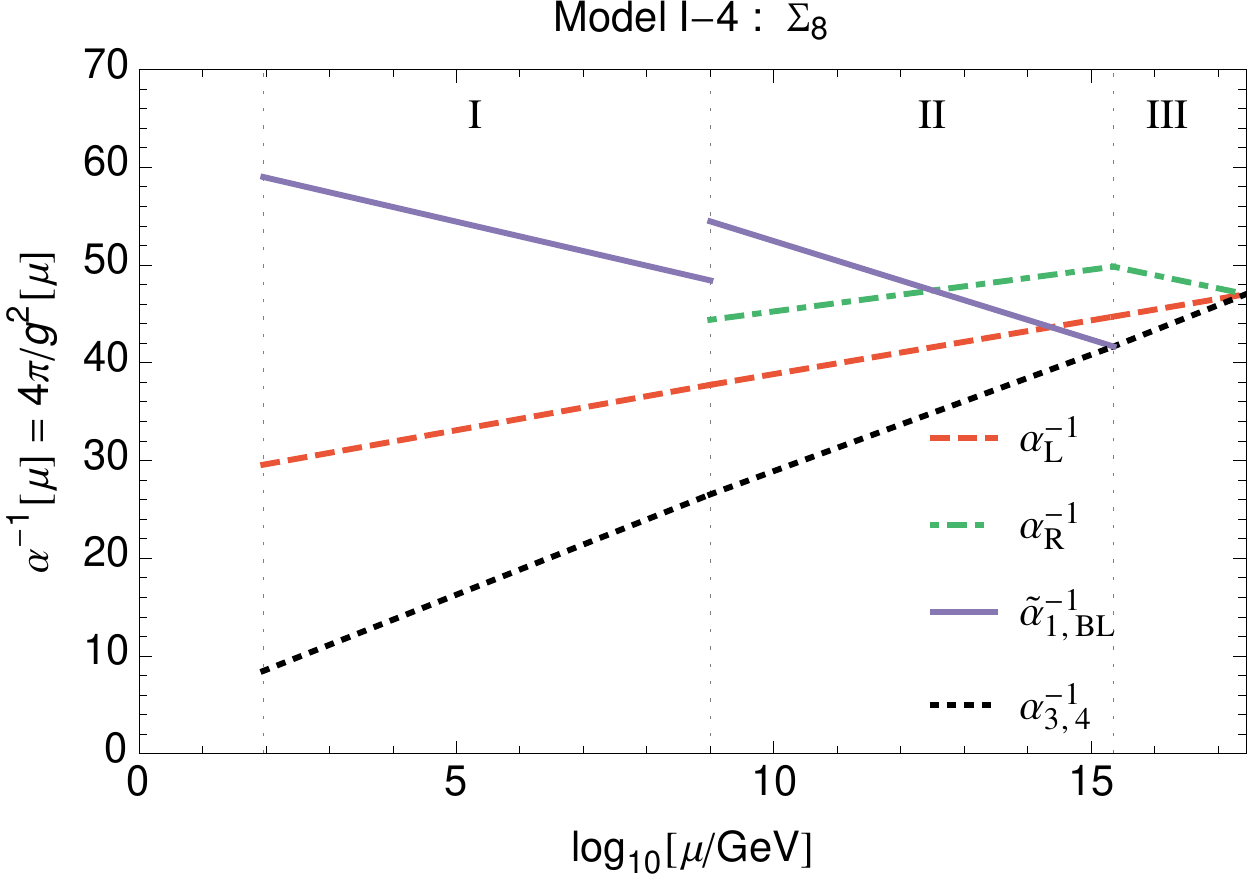}}\hspace{0.4cm}
  \subfigure[$\;\;(r,c,d,u)=(5 \mbox{ TeV},\;14.2,\;14.2,\;19.45)$ ]{\includegraphics[width=6.7cm]{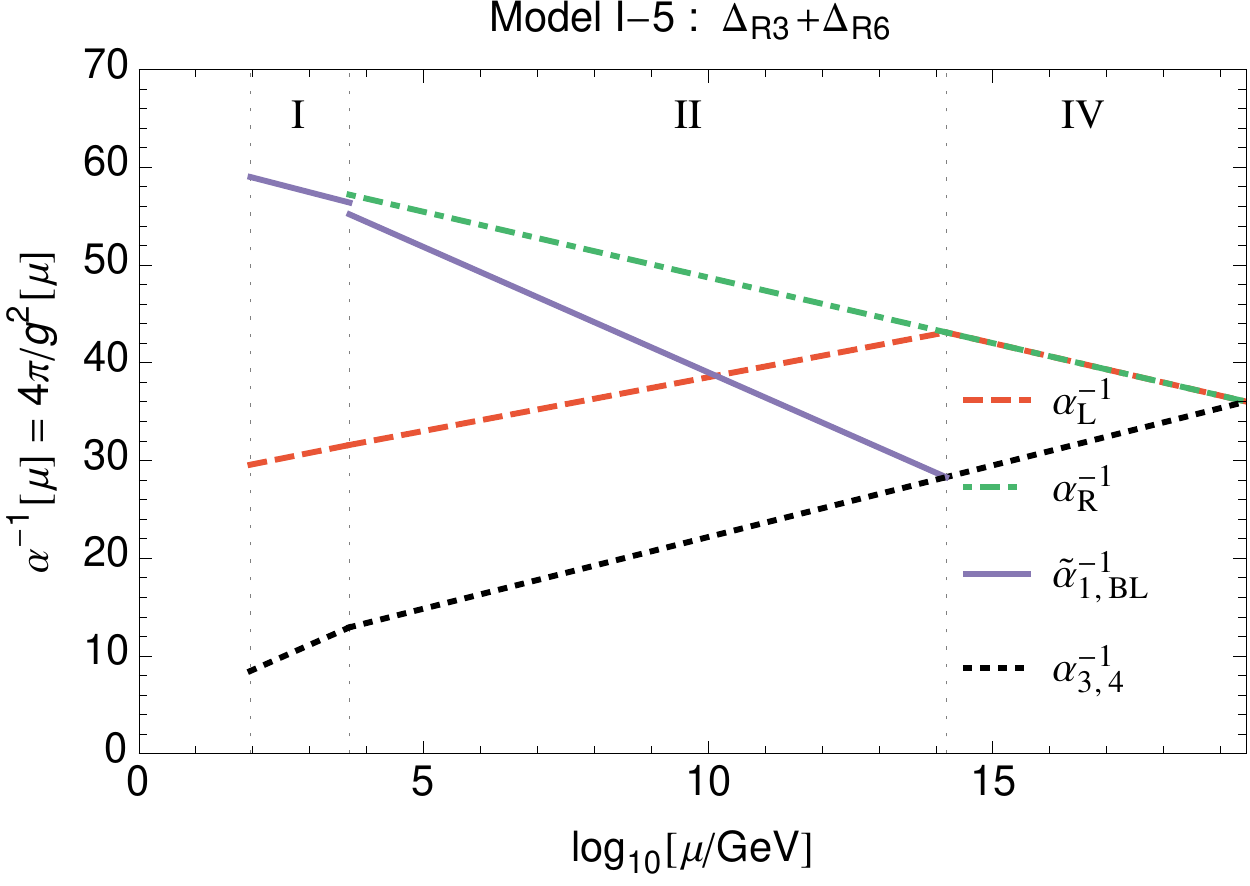}}
    \subfigure[$\;\;(r,c,d,u)=(5 \mbox{ TeV},\;17.5,\;17.5,\;18.7)$ ]{\includegraphics[width=6.7cm]{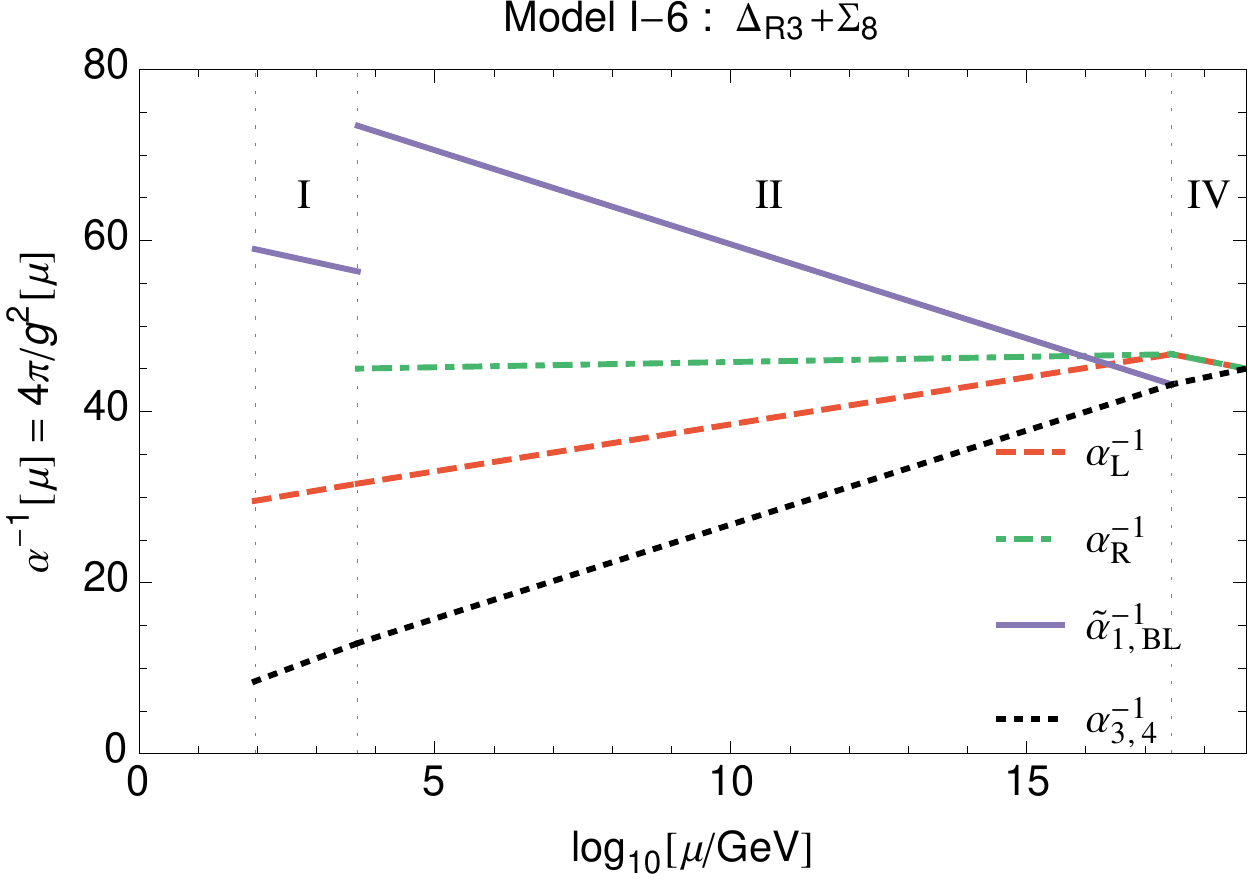}}
\caption{Running of the gauge couplings for Model's I-2 through I-6. Model's I-2, I-5, and I-6 allow $M_R=5$ TeV, and thus $M_R$ is fixed to this value for these  cases. Fixing $M_R$ narrows down the intervals of allowed values for the other scales. Since at this point there is only one unknown left, selecting a value for one of the other scales automatically determines the other two. For Model's I-3 and I-4, we select a $(M_R, M_U)$ combination which gives a better physical interpretation compared to the other combinations. For example, for a particular  $(M_R, M_U)$ pair, if two adjacent scales are numerically too close to each other, the physical interpretation is not clear. Then we chose a combination which separates these values apart or makes them exactly the same.}
\label{RGrunning2}
\end{figure}

The case being such, the system does not allow that $M_R=5$ TeV. The rest of the results are displayed in Table~\ref{valuesI4}. The running of the coupling constants is given in FIG.~\ref{RGrunning2} (c), for a sample of values of the symmetry breaking scales. 

\subsubsection{Model I-5:  A triplet + a sextet}
In this case, we have both the color-triplet ($\Delta_{R3}$) and the color-sextet  ($\Delta_{R6}$) components of the Higgs multiplet $\Delta_{R}(1,3,10)$ in the interval II ($M_R-M_C$), where the gauge group is $G_{2213}$, in addition to our usual Higgs fields. Then, the scalar content in the energy interval II is given as
\begin{equation}
\phi(2,2,0,1)\;,\quad\Delta_{R1}(1,3,2,1)\;,\quad\Delta_{R3}\left(1,3,\frac{2}{3},3\right)\;,\quad\Delta_{R6}\left(1,3,\dfrac{-2}{3},6\right)\;.
\end{equation}
The corresponding RG coefficients for the interval II are given in Table \ref{Higgscontent1} and the ones for the other intervals are given in Table \ref{a1}. Using Eqs.~(\ref{A3}-\ref{A4}), in terms of the definitions given in Eq.~(\ref{xyzDef}), we obtain
\begin{eqnarray}
\label{eq5} 
1418 &=&-46 u+82 d + 84 c-11r\;,\cr
993 &=& 46 u -2d+ 4 c + 19 r\;.
\end{eqnarray}
%


\begin{table}[b]
  \centering 
  \caption{The predictions of Model I-5. In the $M_R$-floating case, we do not allow $M_U$ to exceed the underlined value.}
  \label{values5}
  \begin{tabular}{cc}
  {\begin{tabular}{ccccc}
  \toprule
\ \ \ $\;\;M_X$\ \ \ \ \ & $\qquad $ $\log_{10} M_X/GeV$ $\qquad$ \\
\colrule
$\vphantom{\bigg|}$ $M_U$ &   $[15.3, \;\underline{19.45}]$  \\
$\vphantom{\bigg|}$ $M_D$ &                $[13.7, \;19.45]$  \\
$\vphantom{\bigg|}$ $M_C$ &                $[9.2, \;14.2]$  \\
$\vphantom{\bigg|}$ $M_R$  &  $[3.7,\; 13.7]$   \\ \hline
$\vphantom{\bigg|}$ $\alpha_U^{-1}$ &               $[36.0,\;49.0]$  \\ \hline
$\vphantom{\bigg|}$ $ g_R(M_R) $ &               $[0.43,\;0.54]$  \\
\botrule
\end{tabular}
}\qquad&\qquad
{\begin{tabular}{ccccc}
\toprule
\ \ \ $\;\;M_X$\ \ \ \ \ & $\qquad $ $\log_{10} M_X/GeV$ $\qquad$ \\
\colrule
$\vphantom{\bigg|}$ $M_U$ &   $19.45$  \\
$\vphantom{\bigg|}$ $M_D$ &                $14.2$  \\
$\vphantom{\bigg|}$ $M_C$ &                $14.2$  \\
$\vphantom{\bigg|}$ $M_R$  &  $5$ TeV  \\ \hline
$\vphantom{\bigg|}$ $\alpha_U^{-1}$ &               $36.0$  \\ \hline
$\vphantom{\bigg|}$ $ g_R(M_R) $ &               $0.47$  \\
\botrule
\end{tabular}
}\\
(a) $M_R$-floating.  & (b) $M_R$-fixed.
\end{tabular}
\end{table}

Solving these equations, while maintaining the order of breaking scales, we obtain
\begin{eqnarray}
\label{minbad2}
(M_R)_{min}\;:\qquad M_R\;=\; M_Z\;, \quad  \frac{M_U}{\mbox{GeV}}&=&\left[10^{20.2},\; 10^{21.0}\right]\;,
\quad \frac{M_D}{\mbox{GeV}}\;=\;\left[10^{14.3},\; 10^{21.0}\right]\;\nonumber\\
 \frac{M_C}{\mbox{GeV}}&=&\left[10^{8.1},\; 10^{14.3}\right]\;.
 \end{eqnarray}
Here, we have a similar situation as in Model I-3 that $M_U$ (and partially $M_D$) exceeds the Planck scale. If we employ the condition of maintaining the scales below the Planck mass, we obtain 
\begin{eqnarray}
\label{minbad3}
(M_R)_{min}\;: \quad M_R\;=\; 62\mbox{ TeV}\;, \quad  \frac{M_U}{\mbox{GeV}}&=&10^{19.0}\;,\quad  \frac{M_D}{\mbox{GeV}}\;=\;\frac{M_C}{\mbox{GeV}}\;=\; 10^{14.1}\;,
\end{eqnarray}
where $(M_R)_{min}$ is above the TeV scale. 

What is different in this case is that if we slightly relax our constraint on $(M_U)_{max}$, we obtain a TeV scale $M_R$ where $(M_U)_{max}=2.8\times 10^{19}$ GeV, which is only slightly above the Planck mass. If we set this new value as the upper bound, we find
\begin{eqnarray}
\label{minI5}
(M_R)_{min}\;:\quad M_R\;=\; 5\mbox{ TeV}\;, \quad  \frac{M_U}{\mbox{GeV}}&=&10^{19.45}\;,\quad  \frac{M_D}{\mbox{GeV}}\;=\;\frac{M_C}{\mbox{GeV}}\;=\; 10^{14.2}\;.
\end{eqnarray}

The rest of the results for the final case are displayed in Table~\ref{values5}, and the running of the coupling constants is given in FIG.~\ref{RGrunning2} (d).

\subsubsection{Model I-6:  A triplet + an octet}

In this model, we have the color-triplet ($\Delta_{R3}$) component of $\Delta_{R}(1,3,10)$ and the color-octet  ($\Sigma_{8}$) component of $\Sigma(1,1,15)$ in the interval II ($M_R-M_C$), where the gauge group is $G_{2213}$, in addition to our usual Higgs fields. The scalar content in the energy interval II becomes
\begin{equation}
\phi(2,2,0,1)\;,\quad\Delta_{R1}(1,3,2,1)\;,\quad\Delta_{R3}\left(1,3,\frac{2}{3},3\right)\;,\quad\Delta_{8}\left(1,1,0,8\right)\;.
\end{equation}
The corresponding RG coefficients for the interval II are given in Table \ref{Higgscontent1} and the ones for the other intervals are given in Table \ref{a1}. Using the Eqs.~(\ref{A3}-\ref{A4}), in terms of the definitions in Eq.~(\ref{xyzDef}), we have
\begin{eqnarray}
\label{eq6} 
1418 &=&-46 u+82 d + 42 c+31r\;,\cr
993 &=& 46 u -2d+ 6 c + 17 r\;.
\end{eqnarray}
%

\begin{table}[t]
  \centering 
  \caption{The predictions of Model I-6. Since the system itself does not maintain $M_U$ below the Planck scale, we externally apply this condition in both $M_R$-floating and $M_R$-fixed cases.}
  \label{valuesa6}
  \begin{tabular}{cc}
  {\begin{tabular}{ccccc}
  \toprule
\ \ \ $\;\;M_X$\ \ \ \ \ & $\qquad $ $\log_{10} M_X/GeV$ $\qquad$ \\
\colrule
$\vphantom{\bigg|}$ $M_U$ &   $[15.3, \;\underline{19.0}]$  \\
$\vphantom{\bigg|}$ $M_D$ &                $[13.7, \;19.0]$  \\
$\vphantom{\bigg|}$ $M_C$ &                $[11.2, \;17.8]$  \\
$\vphantom{\bigg|}$ $M_R$  &   $[2.8, \;13.7]$ \\ \hline
$\vphantom{\bigg|}$ $\alpha_U^{-1}$ &               $[41.0,\;48.4]$  \\ \hline
$\vphantom{\bigg|}$ $ g_R(M_R) $ &               $[0.48,\;0.54]$  \\
\botrule
\end{tabular}
}\qquad&\qquad
{\begin{tabular}{ccccc}
\toprule
\ \ \ $\;\;M_X$\ \ \ \ \ & $\qquad $ $\log_{10} M_X/GeV$ $\qquad$ \\
\colrule
$\vphantom{\bigg|}$ $M_U$ &   $[18.7, \;\underline{19.0}]$  \\
$\vphantom{\bigg|}$ $M_D$ &                $[17.5, \;18.6]$  \\
$\vphantom{\bigg|}$ $M_C$ &                $[15.6, \;17.5]$  \\
$\vphantom{\bigg|}$ $M_R$  &  $5$ TeV  \\ \hline
$\vphantom{\bigg|}$ $\alpha_U^{-1}$ &               $[45.0,\;47.4]$  \\ \hline
$\vphantom{\bigg|}$ $ g_R(M_R) $ &               $[0.50,\;0.53]$  \\
\botrule
\end{tabular}
}\\
(a) $M_R$ floating:  & (b) $M_R$ fixed:
\end{tabular}
\end{table}

Maintaining the order of symmetry breaking scales, we find that $(M_R)_{min}=M_Z$, and when $M_R=(M_R)_{min}$, $M_U/\mbox{GeV}=[10^{19.3},\;10^{19.7}]$, which is slightly above the Planck mass. Imposing that $(M_U)_{max}=M_P$, we find
\begin{eqnarray}
(M_R)_{min}&:&\qquad M_R\;=\; 690 \mbox { GeV}\;,\quad \dfrac{M_U}{\mbox{ GeV}}\;=\; 10^{19.0}\;, \quad \dfrac{M_C}{\mbox{ GeV}}\;=\;\dfrac{M_D}{\mbox{ GeV}}\;=\; 10^{17.8}\;.\nonumber\\
\end{eqnarray}
Therefore, the system allows for $M_R=5$ TeV. The results are displayed in Table~\ref{valuesa6}, and the running of the coupling constants is given in FIG.~\ref{RGrunning2} (e).

%

%
%

\subsection{Models type-II}
We define the models type-II as the models whose symmetry breaking sequence is Chain-II, where the ordering of $M_C$ and $M_D$ is reversed, which is given as
\begin{equation}
\mbox{Chain II:} \qquad SO(10) \;\underset{54}{\xrightarrow{M_U}}\; G_{224D} \;\underset{210}{\xrightarrow{M_C}}\; G_{2213D} \;\underset{210}{\xrightarrow{M_D}}\;G_{2213}\;\underset{126}{\xrightarrow{M_R}}\; G_{213} \;\underset{10}{\xrightarrow{M_Z}}\; G_{13}\;.
\end{equation}
The first part of the symmetry breaking is accomplished as before by $\mathbf{54}$ which contains a $G_{224D}$ singlet in its decomposition. In the second stage,  where only the $SU(4)_C$ is broken but the $D$-parity is not, the parity-even field $(1,1,15)_{210}\equiv \Sigma'$ is used. The multiplet $(1,1,15)_{45}$  could be used in the third stage, where only the parity is broken, since it contains the required, parity-odd, $G_{2213}$-singlet field, $(1,1,0,1)_{45}$. However, since in our systematic study we try to keep the high energy Higgs content as minimal as possible, we choose to use the singlet  $\sigma$, contained in $\mathbf{210}$, as we did in the previous section. Note that this is the only other option to break the $D$-parity\footnote{Using $(1,1,0,1)_{45}$ would require to include $(1,1,15)_{45}$ in the interval IV in addition to $(1,1,15)_{210}$. In terms of the RG evolution, this extra multiplet in IV wouldn't change the results in a noticeable manner in any case, because its effect in RG equations would appear as a contribution to the term $(a_L + a_R -2 a_4)$, which would be significantly small compared to the rest of the term.}. The rest of the symmetry breaking proceeds in the same way as before.

We will proceed in the rest of this section as follows. We will first work out the ESH case, where the Higgs content is chosen according to the extended survival hypothesis and show that it does not allow $M_R$ to be in the TeV-scale. After that, as in the previous section, we will look at various scenarios where some of the colored scalars survive down to low energies. Among the latter ones, we will focus only on the working scenarios, by which we refer the ones that allow $M_R$ to be in the TeV-scale.

\subsubsection{Model II-1: ESH}
Under the ESH, the scalar content of this model in the energy intervals I through IV are:
\begin{eqnarray}
\mathrm{IV} & \;:\; & \sigma(1,1,1)\;,\;\phi(2,2,1)\;,\;\Delta_R (1,3,10)\;,\;\Delta_L (3,1,10)\;,\;\Sigma'(1,1,15)\;,\cr
\mathrm{III} & \;:\; & \sigma(1,1,0,1)\;,\;\phi(2,2,0,1)\;,\;\Delta_{R1}(1,3,2,1)\;,\;\Delta_{L1}(3,1,2,1)\;,\cr
\mathrm{II}  & \;:\; & \phi(2,2,0,1)\;,\;\Delta_{R1}(1,3,2,1)\;, \vphantom{\bigg|}\cr
\mathrm{I}   & \;:\; & \phi_2(2,1,1)\;.
\end{eqnarray}
The values of the RG coefficients for this Higgs content for this model are listed in Table~\ref{E1}.
\begin{table}[b]
\caption{The Higgs content and the RG coefficients in the four energy intervals for the Model II-1 where the Higgs selection is made according to the ESH.}
{\begin{tabular}{c|l|l}
\hline
$\vphantom{\Big|}$ Interval & Higgs content & RG coefficients
\\
\hline
$\vphantom{\Biggl|}$   IV
& $\Delta_R (1,3,10),\;\Delta_L (3,1,10),\;\Sigma'(1,1,15),$
& $\left( a_{L},a_{R},a_{4}\right)^\mathrm{IV}
 =\left(\dfrac{11}{3},\dfrac{11}{3},-4\right)$
\\
& $\sigma(1,1,1),\;\phi(2,2,1)$ & \\
& &
\\
\hline
$\vphantom{\Biggl|}$ III
& $\Delta_{R1} (1,3,2,1),\;\Delta_{L1} (3,1,2,1),$
&$\left(a_{L},a_{R},a_{BL},a_3\right)^\mathrm{III}=\left(\dfrac{-7}{3},\dfrac{-7}{3},\dfrac{14}{3},-7\right)$
\\
& $\sigma(1,1,0,1),\;\phi(2,2,0,1)$ & \\
& &
\\
\hline
$\vphantom{\Biggl|}$  II
& $\phi(2,2,0,1),\;\Delta_{R1}(1,3,2,1)$
& $\left(a_{L},a_{R},a_{BL},a_3\right)^\mathrm{II}=\left(-3,\dfrac{-7}{3},\dfrac{11}{3},-7\right)$
\\
\hline
$\vphantom{\Biggl|}$   I
& $\phi_2(2,1,1)$
& $\left(a_{1},a_{2},a_{3}\right)^\mathrm{I}=\left(\dfrac{41}{6},\dfrac{-19}{6},-7\right)$
\\
\hline
\end{tabular}}
\label{E1}
\end{table}
Using the Eqs.~(\ref{A8}) and (\ref{A9}) together with the values in Table \ref{E1}, we obtain
\begin{eqnarray}
\label{eqscalesE1}
2774 &=&
 -46\ln\dfrac{M_U}{M_C}
+56\ln\dfrac{M_C}{M_D}
+57\ln\dfrac{M_D}{M_R}
+109\ln\dfrac{M_R}{M_Z}
\;,\cr
1985 & = &
46\ln\dfrac{M_U}{M_C}
+56\ln\dfrac{M_C}{M_D}
+51\ln\dfrac{M_D}{M_R}
+67\ln\dfrac{M_R}{M_Z}
\;.
\end{eqnarray}

In terms of the parameters defined in Eq.~(\ref{xyzDef}),  Eq.~(\ref{eqscalesE1}) becomes
\begin{eqnarray}
\label{eqE1} 
1418 &=&-46 u+102c + d+ 52r\;,\;\\
 993 &=& 46 u+10c- 5 d + 16 r\;.
\end{eqnarray}
Now, the constraint we should take into account for this model, which is the second relation in Eq.~(\ref{ordering}), is given in terms of these parameters as
\begin{equation}
\label{ordering2}
\ln M_Z \;\le\; r \;\le\; d \;\le\; c \;\le\; u\;.
\end{equation}
Numerically solving Eq.~(\ref{eqE1}) numerically we obtain the minimum value allowed for $r$ in this model when $u=c=d$, the maximum value when $r=d=c$. Since in both cases, the ordering between $d$ and $c$ does not apply, the situation is exactly the same as in Model I-1, which is given in Eqs.~(\ref{min1}) and (\ref{max1}).  Since the minimum allowed value for $M_R$ in this model is $M_R= 10^{9.03}$ GeV, it does not serve for our purpose of obtaining a TeV-scale $M_R$.
\begin{center}
\begin{table}[t]
\caption{The predictions of Model II-1.}
{\begin{tabular}{ccccc}
\toprule
\ \ \ $\;\;M_X$\ \ \ \ \ & $\qquad $ $\log_{10} M_X/GeV$ $\qquad$ \\
\colrule
$\vphantom{\bigg|}$ $M_U$ &   $[15.3, \;16.6]$  \\
$\vphantom{\bigg|}$ $M_C$ &                $[13.7, \;16.6]$  \\
$\vphantom{\bigg|}$ $M_D$ &                $[10.1, \;16.6]$  \\
$\vphantom{\bigg|}$ $M_R$ &  $[9.0, \;13.7]$   \\ \hline
$\vphantom{\bigg|}$ $\alpha_U^{-1}$ &               $[41.0,\;46.2]$  \\
\botrule
\end{tabular}
\label{valuesE1}}
\end{table}
\end{center}
\vspace{-1.3cm}

 All the other ranges of values predicted in this model are summarized in Table \ref{valuesE1}. Note that some other boundary values also are exactly the same as the ones in Model I-1, given in Table \ref{values1}. This is again because the conditions for getting these boundary values in those intervals involve the sub-condition $d=c$, which removes the effect of ordering between $d$ and $c$ (and thus between $M_C$ and $M_D$), as in the case of finding the boundary values for $M_R$, which is explained above.

The running of the coupling constants for this case is given in FIG.~\ref{RGrunning3} (a), for a sample of values for the symmetry breaking scales. 

\subsubsection{Model II-2: A triplet + a sextet}
In this case, we investigate the scenario in which there is a light triplet ($\Delta_{R3}$) and a light sextet ($\Delta_{R6}$) which survive in the RG equations down to $M_R$, in addition to the usual light Higgs content, $\Delta_{R1}$ and $\phi$. Therefore, in this model, the members of the multiplet $\Delta_R(1,3,10)$ pick light masses altogether. Then, the Higgs content in the intervals III and II are
\begin{eqnarray}
\mathrm{III} & \;:\; & \sigma(1,1,0,1)\;,\;\phi(2,2,0,1)\;,\;\Delta_{R1}(1,3,2,1)\;,\;\Delta_{L1}(3,1,2,1)\;,\nonumber\\
\vphantom{\Biggl|}
&& \Delta_{R3}\left(1,3,\dfrac{2}{3},3\right),\;\Delta_{L3}\left(3,1,\dfrac{2}{3},3\right),\;\Delta_{R6}\left(1,3,\dfrac{-2}{3},6\right),\;\Delta_{L6}\left(3,1,\dfrac{-2}{3},6\right),\nonumber\\
\mathrm{II}  & \;:\; & \phi(2,2,0,1)\;,\;\Delta_{R1}(1,3,2,1)\;,\Delta_{R3}\left(1,3,\dfrac{2}{3},3\right),\;\Delta_{R6}\left(1,3,\dfrac{-2}{3},6\right)\;.
\end{eqnarray}

Note that in the interval III we take into account $\Delta_L$ components as well, since the relevant symmetry group is $G_{2213D}$. The values of the RG coefficients for the Higgs content in the intervals II and III are listed in Table~\ref{E2&E3}. Since the intervals I and IV are unchanged from the previous model, the RG coefficients for these intervals are the same as the ones given in Table \ref{E1}. 

Using Eqs.~(\ref{A8}) and (\ref{A9}) together with the relevant RG coefficients, in terms of the definitions given in Eq.~(\ref{xyzDef}), we obtain
\begin{eqnarray}
\label{eqE2} 
1418 &=&-46 u+81c +85 d-11r\;,\;\\
 993 &=& 46 u+7c- 5 d + 19 r\;.
\end{eqnarray}
%

\begin{table}[t]
\caption{The Higgs content and the RG coefficients in the intervals III and II for Model II-2 and Model II-3. No change in the intervals I and IV from the Model II-1.}
{\begin{tabular}{c|l|l|l}
\hline
$\vphantom{\Big|}$ Models & Interval & Higgs content & $\left(a_{L},a_{R},a_{BL},a_3\right)$
\\
\hline
$\vphantom{\Biggl|}$  E2& III
& $\Delta_{R1},\;\Delta_{L1},\;\Delta_{R3},\;\Delta_{L3},\;\Delta_{R6},\;\Delta_{L6},\;\phi,\;\sigma$
&$\left(\dfrac{11}{3},\dfrac{11}{3},\dfrac{19}{3},\dfrac{-3}{2}\right)$
\\
\cline{2-4}
$\vphantom{\Biggl|}$ & II
& $\Delta_{R1},\;\Delta_{R3},\;\Delta_{R6},\;\phi$
&$\left(-3,\dfrac{11}{3},\dfrac{14}{3},-4\right)$
\\
\hline
$\vphantom{\Biggl|}$  E3& III
& $\Delta_{R1},\;\Delta_{L1},\;\Delta_{R3},\;\Delta_{L3},\;\Delta_{R6},\;\Delta_{L6},\;\phi,\;\sigma$
& $\left(\dfrac{11}{3},\dfrac{11}{3},\dfrac{19}{3},\dfrac{-3}{2}\right)$
\\
\cline{2-4}
$\vphantom{\Biggl|}$ & II
& $\Delta_{R1},\;\Delta_{R6},\;\phi$
&$\left(-3,\dfrac{5}{3},\dfrac{13}{3},\dfrac{-9}{2}\right)$
\\
\hline
\end{tabular}}
\label{E2&E3}
\end{table}


\begin{table}[t]
  \centering 
  \caption{The predictions of Model II-2. Since the system in this model, unlike the previous one, does not maintain $M_U$ below the Planck scale itself, we externally apply this condition in both $M_R$-floating and $M_R$-fixed cases.}
  \label{valuese2}
  \begin{tabular}{cc}
  {\begin{tabular}{ccccc}
  \toprule
\ \ \ $\;\;M_X$\ \ \ \ \ & $\qquad $ $\log_{10} M_X/GeV$ $\qquad$ \\
\colrule
$\vphantom{\bigg|}$ $M_U$ &   $[15.3, \;\underline{19.0}]$  \\
$\vphantom{\bigg|}$ $M_C$ &                $[13.7, \;18.9]$  \\
$\vphantom{\bigg|}$ $M_D$ &                $[9.2, \;14.1]$  \\
$\vphantom{\bigg|}$ $M_R$  &   $[M_Z, \;13.7]$ \\ \hline
$\vphantom{\bigg|}$ $\alpha_U^{-1}$ &               $[24.4,\;41.0]$  \\ \hline
$\vphantom{\bigg|}$ $ g_R(M_R) $ &               $[0.48,\;0.55]$  \\
\botrule
\end{tabular}
}\qquad&\qquad
{\begin{tabular}{ccccc}
\toprule
\ \ \ $\;\;M_X$\ \ \ \ \ & $\qquad $ $\log_{10} M_X/GeV$ $\qquad$ \\
\colrule
$\vphantom{\bigg|}$ $M_U$ &   $[18.3, \;\underline{19.0}]$  \\
$\vphantom{\bigg|}$ $M_C$ &                $[15.8, \;18.3]$  \\
$\vphantom{\bigg|}$ $M_D$ &                $[9.6, \;12.4]$  \\
$\vphantom{\bigg|}$ $M_R$ &  $5$ TeV  \\ \hline
$\vphantom{\bigg|}$ $\alpha_U^{-1}$ &               $[26.4,\;32.2]$  \\ \hline
$\vphantom{\bigg|}$ $ g_R(M_R) $ &               $[0.49,\;0.52]$  \\
\botrule
\end{tabular}
}\\
(a) $M_R$ floating:  & (b) $M_R$ fixed:
\end{tabular}
\end{table}

\begin{figure}
 \subfigure[$\;\;(r,d,c,u)=(12.0,\;14.8,\;14.8,\;15.8)$ ]{\includegraphics[width=7.5cm]{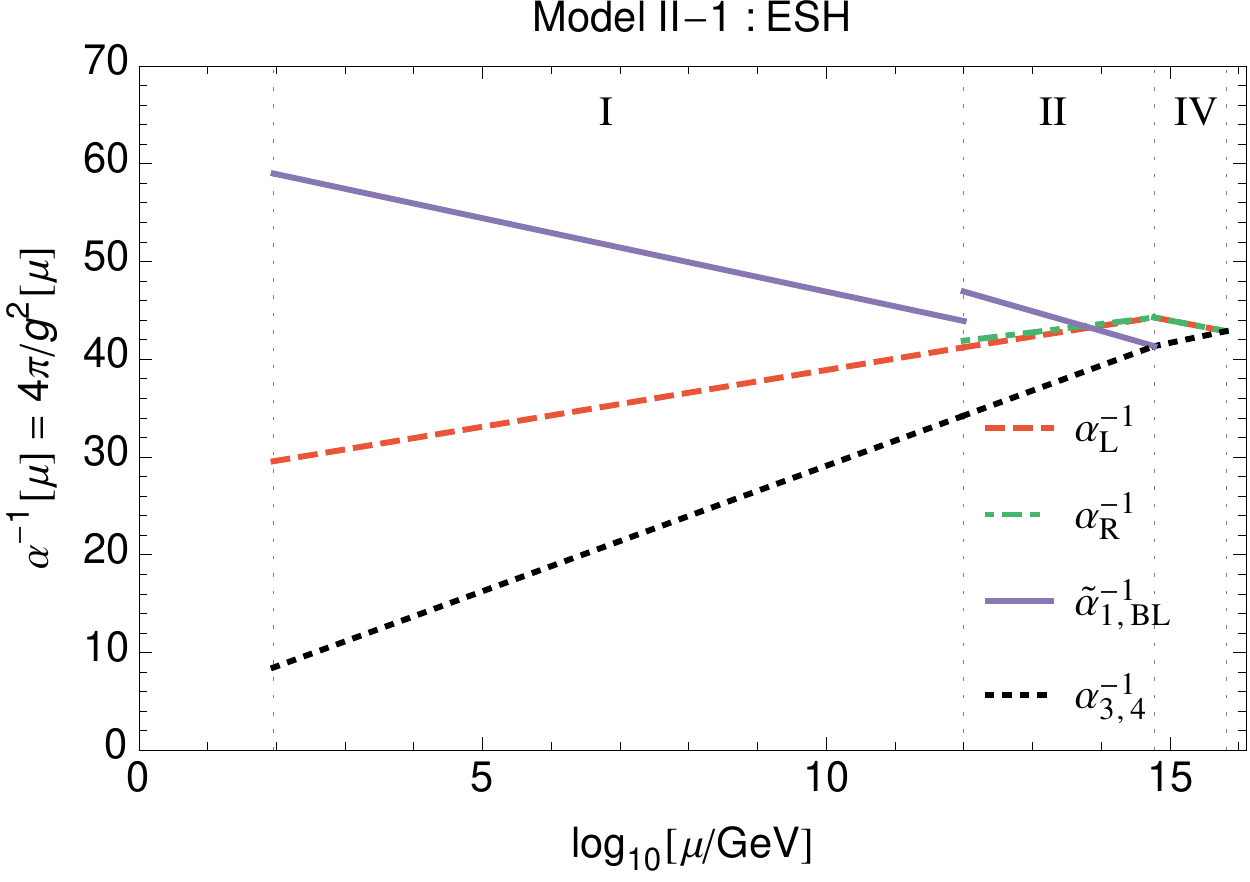}}\hspace{0.4cm}
  \subfigure[$\;\;(r,d,c,u)=(5\mbox{ TeV},\;11.4,\;16.7,\;18.7)$ ]{\includegraphics[width=7.5cm]{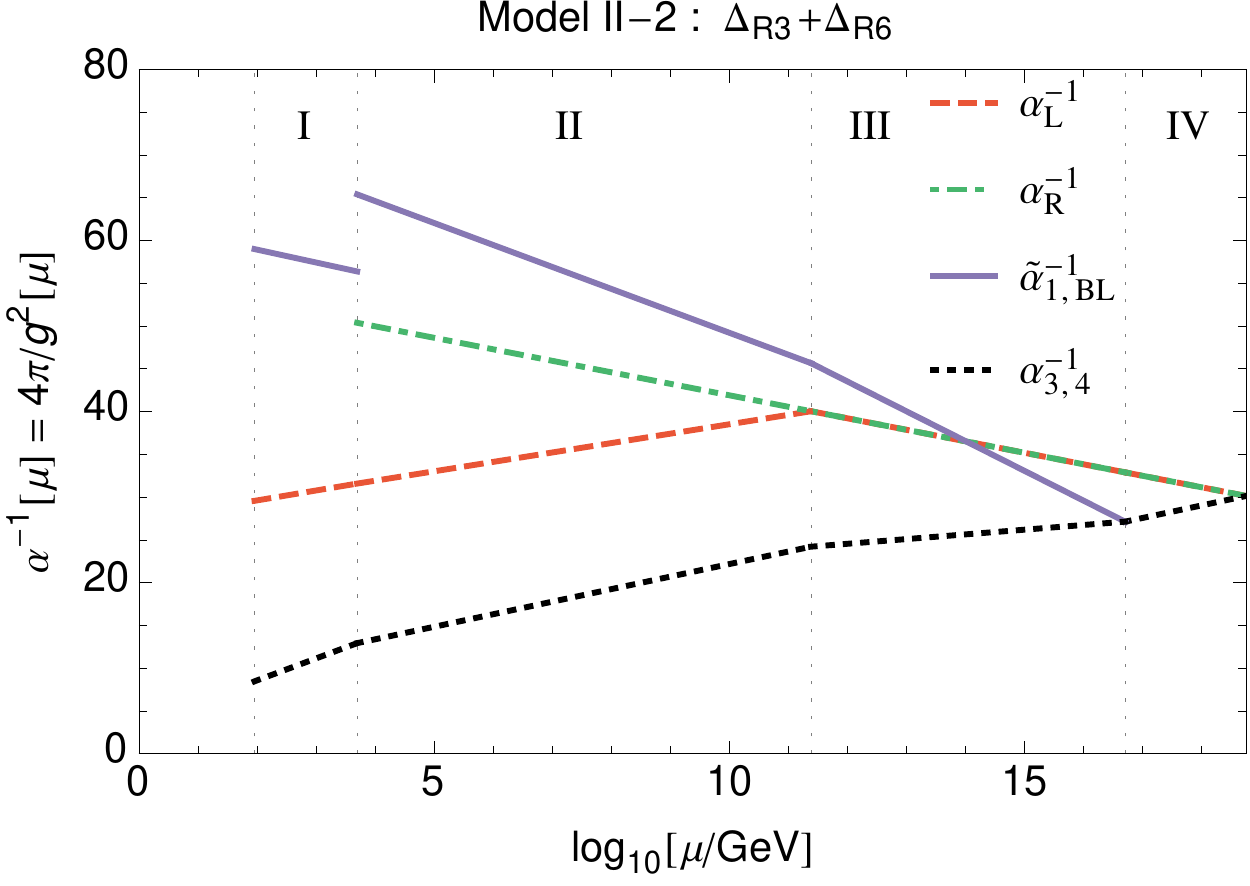}}\vspace{1.0cm}
  \subfigure[$\;\;(r,d,c,u)=(5\mbox{ TeV},\;11.2,\;18.9,\;18.9)$ ]{\includegraphics[width=7.5cm]{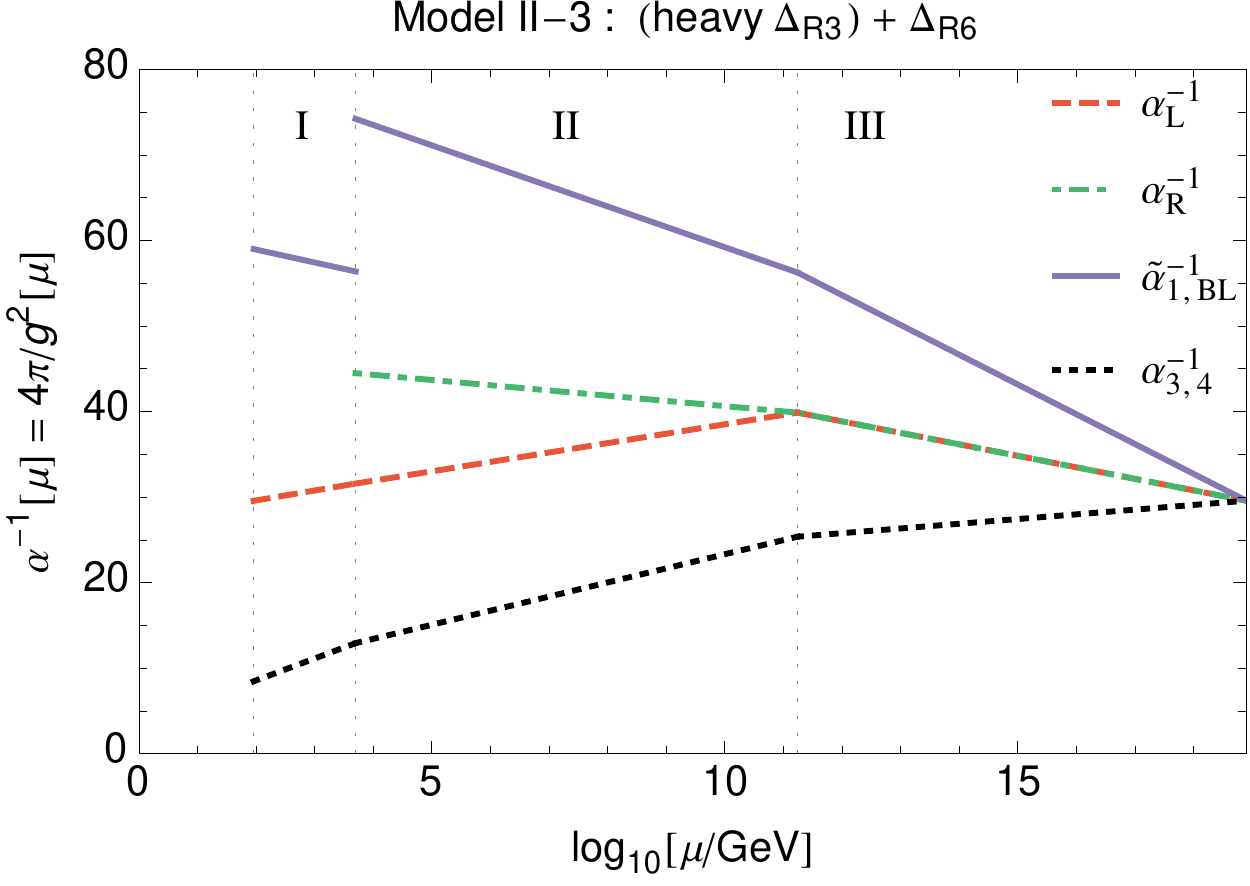}}\hspace{0.4cm}
  \caption{Running of the gauge couplings for the models type II. In (a), Model II-1 is displayed for the a sample of selected values. We prefer to select $M_U$ in such a way that $M_D=M_C$, which collapses the interval III. Note that in the interval II, $\alpha^{-1}_L$ and $\alpha^{-1}_R$  evolve very closely but not identically.
Similarly, in (b) and (c), Model's II-2  and II-3 are displayed. Since these models allow $M_R= 5$ TeV, we fix $M_R$ to this value in these cases.
}
\label{RGrunning3}
\end{figure}

Maintaining the order of symmetry breaking scales, we find that $(M_R)_{min}=M_Z$, and when $M_R=(M_R)_{min}$, $M_U/\mbox{GeV}=[10^{18.9},\;10^{20.2}]$, which is partially above the Planck mass. Imposing that $(M_U)_{max}=M_P$, we find
\begin{eqnarray}
\label{minII2}
(M_R)_{min}\;:\; \qquad M_R\;=\; M_Z\;, \quad  \frac{M_U}{\mbox{GeV}}&=&\left[10^{18.9},\; 10^{19.0}\right]\;,
\quad \frac{M_C}{\mbox{GeV}}\;=\;\left[10^{18.6},\; 10^{18.9}\right]\;\nonumber\\
 \frac{M_D}{\mbox{GeV}}&=&\left[10^{9.2},\; 10^{9.6}\right]\;,
 \end{eqnarray}
whereas the ordered quadruple $(M_U,M_C,M_D,M_R)$ when $M_R=(M_R)_{max}$ has the same pattern of values as the previous models, given in Eq.~(\ref{max1}).
Therefore, the system allows that $M_R= 5$ TeV. The results for which $M_R$ floats and for which it is fixed to $5$ TeV are displayed in Table \ref{valuese2}. The running of the coupling constants is given in FIG.~\ref{RGrunning3} (b).

\subsubsection{Model II-3: Sequential colored-scalars: A heavy triplet and a light sextet}

One of the features of the models type-II compared to the models type-I is that in the former it is possible to have \textit{sequential} colored-scalars. Since the Pati-Salam group is broken to $G_{2213D}$ at $M_C$, a number of scalars may pick their masses of order this scale and decouple from the rest of the system in the RG evolution; a number of them may gain masses of order $M_D$-scale where the D-parity is broken; while the others may have masses in the $M_R$-scale. 

In this final example, we will investigate such a scenario. We will assume the color-triplet $\Delta_{R3}$ gains a mass of order $M_D$ and hence survives in the RG evolution down to this scale, while the mass of the color-sextet $\Delta_{R6}$ is of order $M_R$, and therefore, it survives all the way down to $M_R$. All the others, other than the ones which will acquire VEVs at the subsequent levels of the symmetry breaking, become heavy and decouple in the RG running.

The only difference of this model from the Model II-2 is the Higgs content in the energy interval II, where $\Delta_{R3}$ is absent. The Higgs content in this interval is given as
\begin{equation}
\phi(2,2,0,1)\;,\qquad \Delta_{R1}(1,3,2,1)\;,\qquad \Delta_{R6}\left(1,3,\dfrac{-2}{3},6\right)\;.
\end{equation}
Again, using Eqs.~(\ref{A8}) and (\ref{A9}) together with the corresponding RG coefficients, given in Table \ref{E2&E3}, in terms of the definitions given in Eq.~(\ref{xyzDef}), we obtain following equations.
\begin{eqnarray}
\label{eqE3} 
1418 &=&-46 u+81c +64 d-10r\;,\;\\
 993 &=& 46 u+7c- 8 d + 22r\;.
\end{eqnarray}
%

\begin{table}[b]
  \centering 
  \caption{The predictions of Model II-3. Note that the system itself does not maintain $M_U$ below the Planck scale; therefore, we externally apply this condition.}
  \label{valuese3}
  \begin{tabular}{cc}
  {\begin{tabular}{ccccc}
  \toprule
\ \ \ $\;\;M_X$\ \ \ \ \ & $\qquad $ $\log_{10} M_X/GeV$ $\qquad$ \\
\colrule
$\vphantom{\bigg|}$ $M_U$ &   $[15.3, \;\underline{19.0}]$  \\
$\vphantom{\bigg|}$ $M_C$ &                $[13.7, \;19.0]$  \\
$\vphantom{\bigg|}$ $M_D$ &                $[11.2, \;15.4]$  \\
$\vphantom{\bigg|}$ $M_R$  &   $[3.5, \;13.7]$ \\ \hline
$\vphantom{\bigg|}$ $\alpha_U^{-1}$ &               $[29.5,\;41.0]$  \\ \hline
$\vphantom{\bigg|}$ $ g_R(M_R) $ &               $[0.50,\;0.55]$  \\
\botrule
\end{tabular}
}\qquad&\qquad
{\begin{tabular}{ccccc}
\toprule
\ \ \ $\;\;M_X$\ \ \ \ \ & $\qquad $ $\log_{10} M_X/GeV$ $\qquad$ \\
\colrule
$\vphantom{\bigg|}$ $M_U$ &   $[18.9, \;\underline{19.0}]$  \\
$\vphantom{\bigg|}$ $M_C$ &                $[18.7, \;18.9]$  \\
$\vphantom{\bigg|}$ $M_D$ &                $[11.2, \;11.6]$  \\
$\vphantom{\bigg|}$ $M_R$ &  $5$ TeV  \\ \hline
$\vphantom{\bigg|}$ $\alpha_U^{-1}$ &               $[29.6,\;30.3]$  \\ \hline
$\vphantom{\bigg|}$ $ g_R(M_R) $ &               $\simeq 0.53$  \\
\botrule
\end{tabular}
}\\
(a) $M_R$ floating:  & (b) $M_R$ fixed:
\end{tabular}
\end{table}

Maintaining the order of symmetry breaking scales, we find that $(M_R)_{min}=M_Z$, and when $M_R=(M_R)_{min}$, $M_U/\mbox{GeV}=[10^{19.6},\;10^{21.0}]$, which is above the Planck mass. Imposing that $(M_U)_{max}=M_P$, we find
\begin{eqnarray}
\label{minII3}
(M_R)_{min}\;:\; \qquad M_R\;=\; 2.9 \mbox{ TeV}\;, \quad  \frac{M_U}{\mbox{GeV}}\;=\; \frac{M_C}{\mbox{GeV}}\;=\;10^{19.0}\;,
\quad \frac{M_D}{\mbox{GeV}}\;=\; 10^{11.2}\;,
 \end{eqnarray}
whereas $(M_R)_{max}$ and the corresponding values of the other scales when $M_R=(M_R)_{max}$ are the same as the previous models. Therefore, the system allows that $M_R=5$ TeV. The results\footnote{Note that the color-triplet being heavy is also appealing because of the proton decay; the value of $M_D$ in the $M_R=5$ TeV case, $M_D\simeq 10^{11}$ GeV, coincides with the naive lower bound on the mass of the color-triplet from the limits on the proton decay~\cite{DiLuzio:2011my}. This is an improvement compared to the type-I models, where a suppression mechanism is required for the (light) color-triplet related terms in the proton decay amplitude.} are displayed in Table~\ref{valuese3}, and the running of the coupling constants is given in FIG.~\ref{RGrunning3} (c).

\section{Summary and Conclusions}

In this work, we have addressed the question whether the left-right model, which is based on the gauge group $SU(2)_L\times SU(2)_R\times U(1)_{B-L}\times SU(3)_C$, embedded in the non-supersymmetric $SO(10)$ framework, could explain the recent LHC signals. By performing a detailed renormalization group analysis, we have shown that the regular model, where the Higgs content selection is made under the extended survival hypothesis (ESH), does not allow $M_R$ to be in the TeV-scale. In order to investigate this scheme more in depth, we have relaxed the ESH conjecture and explored the possibility that $M_R$ could be lowered by a number of light colored scalars surviving from $M_C$, at which $SU(4)_C$ is broken, down to $M_R$. We have found that there are several combinations of these colored-scalars, which can serve to this end.  In a few scenarios, the situation is enhanced such that the left-right model is allowed to be in the TeV scale. Note that this may put these colored-scalars within reach of the LHC. However, the predicted values of $g_R(M_R)$ in these models, which lie in the interval $[0.47, \;0.53]$, are not compatible with the range of values $g_R\simeq 0.35-0.45$, given in Ref.~\cite{Brehmer}, required to explain the recent LHC data. 

In this paper, after investigating several models which do not yield positive results for a TeV-scale left-right model, we have only focused on the models which do. However, we have also performed this analysis for other possible combinations of colored scalars in the energy interval II ($M_R-M_C$), including the ones with the other triplets available, $\Sigma_{3}$ and $\Sigma_{\overline{3}}$, which are contained in the decomposition of $\Sigma$, given in Eq.~(\ref{SigmaDeltaphiDecomposition}). The results are very similar to the scenarios discussed in this paper; either these models do not allow $M_R$ to be in the TeV scale; or if they do, the predicted values of $g_R(M_R)$ are similar to the ones obtained in the models we have discussed, with the minimum possible value being $g_R(M_R)\simeq 0.47$. Therefore, we believe that it is not necessary to display them in this paper.

While our analysis could suggest that the left-right model in the $SO(10)$ grand unification scheme is \textit{not} favored by the current LHC data, we note that our results are only valid for the models which have the minimal Higgs content to begin with. Extending the high energy Higgs content may change the outcome. However, this may weaken the predictive power of the scheme, unless there is a strong reasoning behind the modifications made, maintaining the model selection under control.

\section*{Acknowledgments}

We would like to thank Rikard Enberg, Benjamin Ett, Gunnar Ingelman, Tanumoy Mandal, Djordje Minic, and Tatsu Takeuchi for helpful communications and stimulating conversations. This work is supported by the Swedish Research Council under contract 621-2011-5107.

\appendix

\section{Relations Between Symmetry Breaking Scales}
\subsection{Chain I}
For Chain I we have the following relations.
\begin{eqnarray}
\label{A1}
\dfrac{1}{g_1^2(M_Z)}&=&
\dfrac{1}{g_R^2(M_U)} + \dfrac{2}{3}\,\dfrac{1}{g_4^2(M_U)}
+ \dfrac{1}{8\pi^2}\left(a_R + \dfrac{2}{3}a_4\right)^\mathrm{IV}\ln\dfrac{M_U}{M_D}\nonumber\\
+&& \dfrac{1}{8\pi^2}\left(a_R + \dfrac{2}{3}a_4\right)^\mathrm{III}\ln\dfrac{M_D}{M_C}
+ \dfrac{\left(a_R + a_{BL}\right)^\mathrm{II}}{8\pi^2}\ln\dfrac{M_C}{M_R}
+ \dfrac{a_1^\mathrm{I}}{8\pi^2}\ln\dfrac{M_R}{M_Z}
\;,\nonumber\\
\dfrac{1}{g_2^2(M_Z)}& = &
\dfrac{1}{g_L^2(M_U)}
+ \dfrac{a_L^\mathrm{IV}}{8\pi^2}\ln\dfrac{M_U}{M_D}
+ \dfrac{a_L^\mathrm{III}}{8\pi^2}\ln\dfrac{M_D}{M_C}
+ \dfrac{a_L^\mathrm{II}}{8\pi^2}\ln\dfrac{M_C}{M_R}
+ \dfrac{a_2^\mathrm{I}}{8\pi^2}\ln\dfrac{M_R}{M_Z} \;,\nonumber\\
\dfrac{1}{e^2(M_Z)}
& = & \dfrac{1}{g_2^2(M_Z)} + \dfrac{1}{g_1^2(M_Z)}\nonumber\\
&=&
\dfrac{1}{g_L^2(M_U)} + \dfrac{1}{g_R^2(M_U)}+ \dfrac{2}{3}\dfrac{1}{g_{4}^2(M_U)} \nonumber\\
& &
+ \dfrac{1}{8\pi^2}\left(a_L + a_R + \dfrac{2}{3}\,a_4\right)^\mathrm{IV}\ln\dfrac{M_U}{M_D}
+ \dfrac{1}{8\pi^2}\left(a_L + a_R + \dfrac{2}{3}\,a_4\right)^\mathrm{III}\ln\dfrac{M_D}{M_C}\nonumber\\
& &
+ \dfrac{1}{8\pi^2}\left(a_L + a_R + a_{BL}\right)^\mathrm{II}\ln\dfrac{M_C}{M_R}
+ \dfrac{1}{8\pi^2}\left(a_1 + a_2\right)^\mathrm{I}\ln\dfrac{M_R}{M_Z} \;,\nonumber\\
\dfrac{1}{g_3^2(M_Z)}
& = & \dfrac{1}{g_4^2(M_U)}
+ \dfrac{a_4^\mathrm{IV}}{8\pi^2}\ln\dfrac{M_U}{M_D}
+ \dfrac{a_4^\mathrm{III}}{8\pi^2}\ln\dfrac{M_D}{M_C}
+ \dfrac{a_3^\mathrm{II}}{8\pi^2}\ln\dfrac{M_C}{M_R}
+ \dfrac{a_3^\mathrm{I}}{8\pi^2}\ln\dfrac{M_R}{M_Z}
\;,\nonumber\\
\dfrac{1}{g_R^2(M_R)}
& = & \dfrac{1}{g_R^2(M_U)}
+ \dfrac{a_R^\mathrm{IV}}{8\pi^2}\ln\dfrac{M_U}{M_D}
+ \dfrac{a_R^\mathrm{III}}{8\pi^2}\ln\dfrac{M_D}{M_C}
+ \dfrac{a_R^\mathrm{II}}{8\pi^2}\ln\dfrac{M_C}{M_R}
\;.
\end{eqnarray}
If we impose the condition
\begin{equation}
g_L(M_U)\;=\;g_R(M_U)\;=\;g_4(M_U)\;\equiv\; g_U\;,
\end{equation}
then it is straightforward to show that
\begin{eqnarray}
\label{A3}
\lefteqn{
2\pi\left[\dfrac{3-8\sin^2\theta_W(M_Z)}{\alpha(M_Z)}\right]
}\cr
& = &
\Biggl[
 \left(-5a_L+3a_R+2a_4\right)^\mathrm{IV}\ln\dfrac{M_U}{M_D}
+\left(-5a_L+3a_R+2a_4\right)^\mathrm{III}\ln\dfrac{M_D}{M_C}
\cr
& &
+\left(-5a_L + 3a_R + 3a_{BL}\right)^\mathrm{II}\ln\dfrac{M_C}{M_R}
+\left(3a_1 -5a_2\right)^\mathrm{I}\ln\dfrac{M_R}{M_Z}
\Biggr]
\;,
\end{eqnarray}
\begin{eqnarray}
\label{A4}
\lefteqn{
2\pi\left[\dfrac{3}{\alpha(M_Z)} - \dfrac{8}{\alpha_s(M_Z)}\right]
}\cr
& = &
\Biggl[
 \left(3a_L+3a_R-6a_4\right)^\mathrm{IV}\ln\dfrac{M_U}{M_D}
+\left(3a_L+3a_R-6a_4\right)^\mathrm{III}\ln\dfrac{M_D}{M_C}
\cr
& &
+\left(3a_L + 3a_R + 3a_{BL} - 8a_3\right)^\mathrm{II}\ln\dfrac{M_C}{M_R}
+\left(3a_1 + 3a_2 - 8a_3\right)^\mathrm{I}\ln\dfrac{M_R}{M_Z}
\Biggr]
\;,
\end{eqnarray}
\begin{eqnarray}
\label{A5}
\lefteqn{
2\pi\left[\dfrac{4\pi}{g_R^2(M_R)}-\dfrac{\sin^2\theta_W(M_Z)}{\alpha(M_Z)}\right]
} \nonumber\\
& = &
\left[
 \left(a_R - a_L\right)^\mathrm{III}\ln\dfrac{M_D}{M_C}
+\left(a_R - a_L\right)^\mathrm{II}\ln\dfrac{M_C}{M_R}
-a_2^\mathrm{I}\ln\dfrac{M_R}{M_Z}
\right]
\;,
\end{eqnarray}
\begin{eqnarray}
\label{A6}
\dfrac{8\pi^2}{g_U^2}
& = & \dfrac{3}{8}
\Biggl[
\dfrac{2\pi}{\alpha(M_Z)}
-\Biggl\{
  \left(a_L + a_R + \dfrac{2}{3}\,a_4\right)^\mathrm{IV}\ln\dfrac{M_U}{M_D}
+ \left(a_L + a_R + \dfrac{2}{3}\,a_4\right)^\mathrm{III}\ln\dfrac{M_D}{M_C}
\cr
& & \qquad
+ \left(a_L + a_R + a_{BL}\right)^\mathrm{II}\ln\dfrac{M_C}{M_R}
+ \left(a_1 + a_2\right)^\mathrm{I}\ln\dfrac{M_R}{M_Z}
\Biggr\}
\Biggr]
\cr
& = & \dfrac{2\pi}{\alpha_s(M_Z)}
-\left(
  a_4^\mathrm{IV}\;\ln\dfrac{M_U}{M_D}
+ a_4^\mathrm{III}\;\ln\dfrac{M_D}{M_C}
+ a_3^\mathrm{II}\;\ln\dfrac{M_C}{M_R}
+ a_3^\mathrm{I}\;\ln\dfrac{M_R}{M_Z}
\right)
\;.
\end{eqnarray}
Note that $a_L^\mathrm{IV}=a_R^\mathrm{IV}$ since parity is not broken in energy interval IV.

The corresponding relations for Chain Ia, Ib, and Ic can be obtained from those of Chain I simply putting, $M_U=M_D$, $M_D=M_C$, and $M_U=M_D=M_C$, respectively, in the equations above.

\subsection{Chain II}

In Chain II, the ordering of $M_D$ and $M_C$ is reversed but one just \textit{can't} simply obtain the relevant relations between scales by just putting $M_D\leftrightarrow M_C$ in those of Chain I, simply because when the ordering changes the corresponding groups in the relevant intervals change as well (see Eq.~(\ref{IntervalNumber})).
The running equations of the couplings for Chain II are given as

\begin{eqnarray}
\label{A7}
\dfrac{1}{g_1^2(M_Z)}&=&
\dfrac{1}{g_R^2(M_U)} + \dfrac{2}{3}\,\dfrac{1}{g_4^2(M_U)}
+ \dfrac{1}{8\pi^2}\left(a_R + \dfrac{2}{3}a_4\right)^\mathrm{IV}\ln\dfrac{M_U}{M_C}\nonumber\\
+&&\dfrac{\left(a_R + a_{BL}\right)^\mathrm{III}}{8\pi^2}\ln\dfrac{M_C}{M_D}
+ \dfrac{\left(a_R + a_{BL}\right)^\mathrm{II}}{8\pi^2}\ln\dfrac{M_D}{M_R}
+ \dfrac{a_1^\mathrm{I}}{8\pi^2}\ln\dfrac{M_R}{M_Z}
\;,\nonumber\\
\dfrac{1}{g_2^2(M_Z)}& = &
\dfrac{1}{g_L^2(M_U)}
+ \dfrac{a_L^\mathrm{IV}}{8\pi^2}\ln\dfrac{M_U}{M_C}
+ \dfrac{a_L^\mathrm{III}}{8\pi^2}\ln\dfrac{M_C}{M_D}
+ \dfrac{a_L^\mathrm{II}}{8\pi^2}\ln\dfrac{M_D}{M_R}
+ \dfrac{a_2^\mathrm{I}}{8\pi^2}\ln\dfrac{M_R}{M_Z} \;,\nonumber\\
\dfrac{1}{e^2(M_Z)}
& = & \dfrac{1}{g_2^2(M_Z)} + \dfrac{1}{g_1^2(M_Z)}\nonumber\\
&=&
\dfrac{1}{g_L^2(M_U)} + \dfrac{1}{g_R^2(M_U)}+ \dfrac{2}{3}\dfrac{1}{g_{4}^2(M_U)} \nonumber\\
& &
+ \dfrac{1}{8\pi^2}\left(a_L + a_R + \dfrac{2}{3}\,a_4\right)^\mathrm{IV}\ln\dfrac{M_U}{M_C}
+ \dfrac{1}{8\pi^2}\left(a_L + a_R + a_{BL}\right)^\mathrm{III}\ln\dfrac{M_C}{M_D}\nonumber\\
& &
+ \dfrac{1}{8\pi^2}\left(a_L + a_R + a_{BL}\right)^\mathrm{II}\ln\dfrac{M_D}{M_R}
+ \dfrac{1}{8\pi^2}\left(a_1 + a_2\right)^\mathrm{I}\ln\dfrac{M_R}{M_Z} \;,\nonumber\\
\dfrac{1}{g_3^2(M_Z)}
& = & \dfrac{1}{g_4^2(M_U)}
+ \dfrac{a_4^\mathrm{IV}}{8\pi^2}\ln\dfrac{M_U}{M_C}
+ \dfrac{a_4^\mathrm{III}}{8\pi^2}\ln\dfrac{M_C}{M_D}
+ \dfrac{a_3^\mathrm{II}}{8\pi^2}\ln\dfrac{M_D}{M_R}
+ \dfrac{a_3^\mathrm{I}}{8\pi^2}\ln\dfrac{M_R}{M_Z}
\;,\nonumber\\
\dfrac{1}{g_R^2(M_R)}
& = & \dfrac{1}{g_R^2(M_U)}
+ \dfrac{a_R^\mathrm{IV}}{8\pi^2}\ln\dfrac{M_U}{M_C}
+ \dfrac{a_R^\mathrm{III}}{8\pi^2}\ln\dfrac{M_C}{M_D}
+ \dfrac{a_R^\mathrm{II}}{8\pi^2}\ln\dfrac{M_D}{M_R}
\;.
\end{eqnarray}
Then, the relations between scales for Chain II become
\begin{eqnarray}
\label{A8}
\lefteqn{
2\pi\left[\dfrac{3-8\sin^2\theta_W(M_Z)}{\alpha(M_Z)}\right]
}\cr
& = &
\Biggl[
 \left(-5a_L+3a_R+2a_4\right)^\mathrm{IV}\ln\dfrac{M_U}{M_C}
+\left(-5a_L + 3a_R + 3a_{BL}\right)^\mathrm{III}\ln\dfrac{M_C}{M_D}
\cr
& &
+\left(-5a_L + 3a_R + 3a_{BL}\right)^\mathrm{II}\ln\dfrac{M_D}{M_R}
+\left(3a_1 -5a_2\right)^\mathrm{I}\ln\dfrac{M_R}{M_Z}
\Biggr]
\;,
\end{eqnarray}
\begin{eqnarray}
\label{A9}
\lefteqn{
2\pi\left[\dfrac{3}{\alpha(M_Z)} - \dfrac{8}{\alpha_s(M_Z)}\right]
}\cr
& = &
\Biggl[
 \left(3a_L+3a_R-6a_4\right)^\mathrm{IV}\ln\dfrac{M_U}{M_C}
+\left(3a_L + 3a_R + 3a_{BL} - 8a_3\right)^\mathrm{III}\ln\dfrac{M_C}{M_D}
\cr
& &
+\left(3a_L + 3a_R + 3a_{BL} - 8a_3\right)^\mathrm{II}\ln\dfrac{M_D}{M_R}
+\left(3a_1 + 3a_2 - 8a_3\right)^\mathrm{I}\ln\dfrac{M_R}{M_Z}
\Biggr]
\;,
\end{eqnarray}
\begin{eqnarray}
\label{A10}
2\pi\left[\dfrac{4\pi}{g_R^2(M_R)}-\dfrac{\sin^2\theta_W(M_Z)}{\alpha(M_Z)}\right]
=
\left[
\left(a_R - a_L\right)^\mathrm{II}\ln\dfrac{M_D}{M_R}
-a_2^\mathrm{I}\ln\dfrac{M_R}{M_Z}
\right]
\;,
\end{eqnarray}
\begin{eqnarray}
\label{A11}
\dfrac{8\pi^2}{g_U^2}
& = & \dfrac{3}{8}
\Biggl[
\dfrac{2\pi}{\alpha(M_Z)}
-\Biggl\{
  \left(a_L + a_R + \dfrac{2}{3}\,a_4\right)^\mathrm{IV}\ln\dfrac{M_U}{M_C}
+  \left(a_L + a_R + a_{BL}\right)^\mathrm{III}\ln\dfrac{M_C}{M_D}
\cr
& & \qquad
+ \left(a_L + a_R + a_{BL}\right)^\mathrm{II}\ln\dfrac{M_D}{M_R}
+ \left(a_1 + a_2\right)^\mathrm{I}\ln\dfrac{M_R}{M_Z}
\Biggr\}
\Biggr]
\cr
& = & \dfrac{2\pi}{\alpha_s(M_Z)}
-\left(
  a_4^\mathrm{IV}\;\ln\dfrac{M_U}{M_C}
+ a_3^\mathrm{III}\;\ln\dfrac{M_C}{M_D}
+ a_3^\mathrm{II}\;\ln\dfrac{M_D}{M_R}
+ a_3^\mathrm{I}\;\ln\dfrac{M_R}{M_Z}
\right)
\;.
\end{eqnarray}

Note that for Chain II $a_L^\mathrm{IV}=a_R^\mathrm{IV}$ and $a_L^\mathrm{III}=a_R^\mathrm{III}$ since parity is not broken in energy intervals IV and III.

The corresponding relations for Chain IIa can be obtained from those of Chain II by putting $M_U=M_C$.

\providecommand{\href}[2]{#2}\begingroup\raggedright

\end{document}